\documentclass[manuscript]{acmart}

\setcopyright{none}

\usepackage{graphicx}
\usepackage{subcaption}
\usepackage{tabularx}
\usepackage{lipsum}

\usepackage{xcolor}

\usepackage{amssymb}
\usepackage{amsmath}

\usepackage{subcaption}
\usepackage{multirow}
\usepackage{enumitem}
\usepackage{booktabs}

\usepackage{pdfpages} 

\usepackage{booktabs}
\usepackage{array}

\usepackage{eso-pic}
\AddToShipoutPictureBG{%
  \AtPageUpperLeft{%
    \raisebox{-1.8cm}{
      \makebox[\paperwidth][c]{
        \color{gray} Accepted for publication in International Journal of Human-Computer Studies (IJHCS)}
      }%
    }%
  }%

\newcommand{\major}[1]{\textcolor{black}{#1}}
\newcommand{\revise}[1]{\textcolor{black}{#1}}

\begin{document}

\title{Understanding Trust Toward Human versus AI-generated Health Information through Behavioral and Physiological Sensing}


\author{Xin Sun}
\affiliation{
  \institution{University of Amsterdam}
  \country{The Netherlands}
}

\author{Rongjun Ma}
\affiliation{
  \institution{Aalto University}
  \country{Finland}
}

\author{Shu Wei}
\affiliation{
  \institution{Yale University}
  \country{United States}
}

\author{Pablo Cesar}
\affiliation{
  \institution{Centrum Wiskunde \& Informatica (CWI) and Delft University of Technology}
  \country{The Netherlands}
}

\author{Jos A. Bosch}
\affiliation{
  \institution{University of Amsterdam}
  \country{The Netherlands}
}

\author{Abdallah El Ali}
\affiliation{
  \institution{Centrum Wiskunde \& Informatica (CWI) and Utrecht University}
  \country{The Netherlands}
}



\begin{abstract}
As AI-generated health information proliferates online and becomes increasingly indistinguishable from human-sourced information, it becomes critical to understand how people trust and label such content, especially when the information is inaccurate. 
We conducted two \major{complementary studies: (1) a mixed-methods survey (N=142) employing a 2 (source: Human vs. LLM) × 2 (label: Human vs. AI) × 3 (type: General, Symptom, Treatment) design, and (2) a within-subjects lab study (N=40) incorporating eye-tracking and physiological sensing (ECG, EDA, skin temperature). 
Participants were presented with health information varying by source-label combinations and asked to rate their trust, while their gaze behavior and physiological signals were recorded.}
We found that LLM-generated information was trusted more than human-generated content, whereas information labeled as human was trusted more than that labeled as AI. 
Trust remained consistent across information types. Eye-tracking and physiological responses varied significantly by source and label. 
\major{Machine learning models trained on these behavioral and physiological features predicted binary self-reported trust levels with 73\% accuracy and information source with 65\% accuracy.}
Our findings demonstrate that adding transparency labels to online health information modulates trust. Behavioral and physiological features show potential to verify trust perceptions and indicate if additional transparency is needed.
\end{abstract}

\begin{teaserfigure}
  \includegraphics[width=\textwidth]{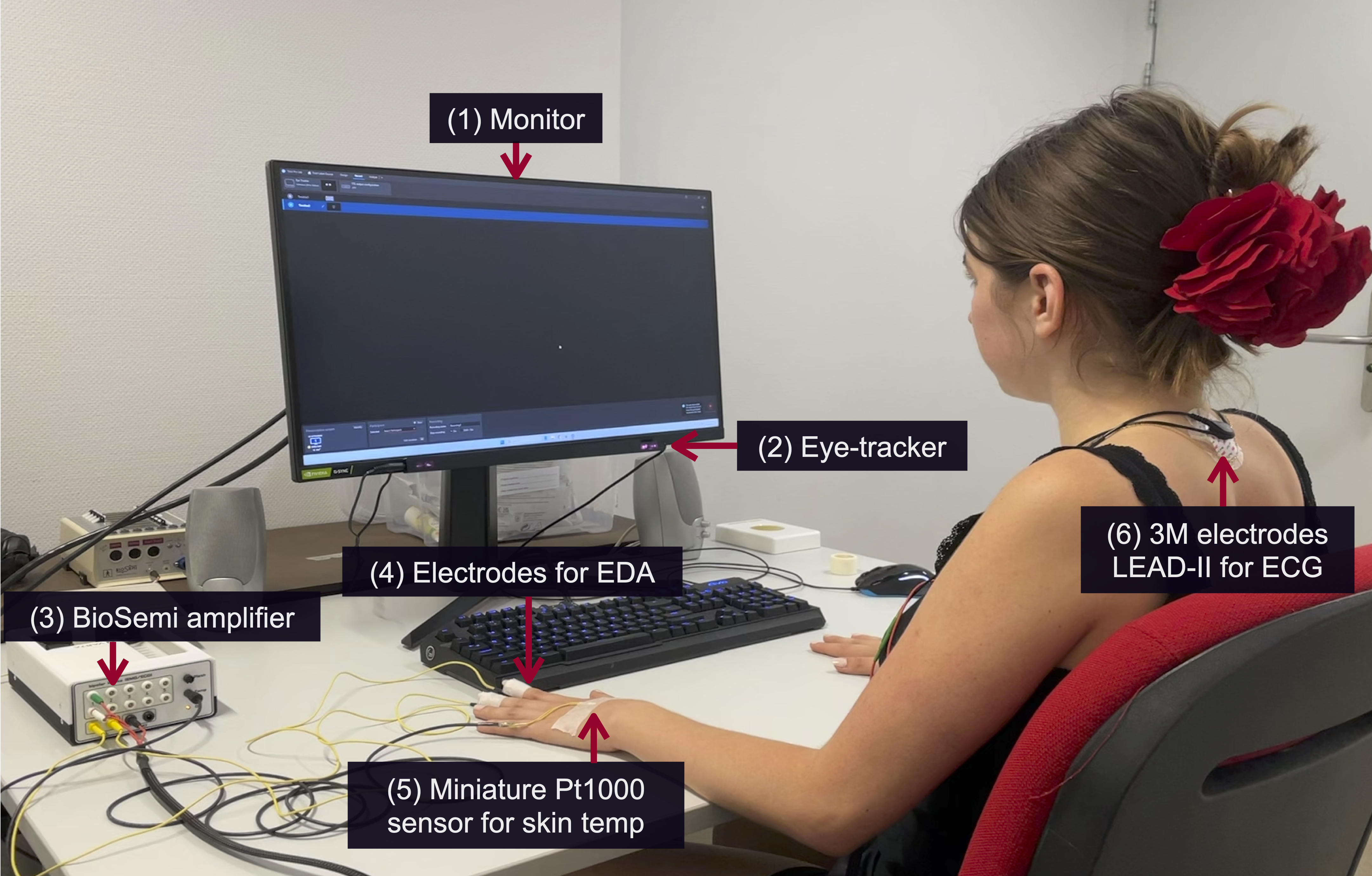}
  \caption{The hardware setup for presenting the text stimulus and collecting physiological signals, eye movement, and pupil dilation.}
  \label{fig:teaser}
\end{teaserfigure}


\maketitle


\section{Introduction}

The internet has become a primary source of health information \cite{Cline, SILLENCE20071853}, with 58.5\% of American adults~\cite{cdcProductsData} (survey on 2022) and 55\% of Europeans~\cite{eurostat} (survey on 2022) using online sources for health-related searches. This shift has transformed how individuals access and engage with health-related content. 
Online health resources encompass a broad range of digital tools,
including professional medical websites~\cite{nih, mayo} and AI-driven tools like health chatbots powered by Large Language Models (LLMs)~\cite{llm}. 
\major{These tools have made health information more accessible and convenient than ever, yet they also require users to make critical choices on which sources of the retrieved health information to trust}~\cite{Liu1, Sillence3}.
These trust decisions directly influence health-related choices, many of which carry significant health risks~\cite{Ethical, journalmedia5020046}. 
As a result, understanding how different information sources shape trust perceptions has become increasingly critical~\cite{bates2006effect}.
\major{Some prior studies find that users tend to trust human-generated information more~\cite{Broom2005, Kerstan, Walker, Reis_Moritz}, while other work suggests that people may prefer algorithmic or AI over human judgments~\cite{Logg,shekar2024peopletrustaigeneratedmedical}.
These mixed findings suggest that trust in online information varies by source and context, and remains insufficiently understood, especially in the LLM-powered health contexts.
}

\major{Disclosed labeling of online information signals its source, but can also shape perceptions independently of the actual source, making it an essential dimension of understanding trust.} 
Misleading labels or unclear sourcing may result in misinformation and poor health decisions~\cite{misinformation,journalmedia5020046}.
Labeling is increasingly mandated by regulations, such as the European AI Act~\cite{transparent_ai}. 
Research shows that disclosed labeling (e.g., with/without indicating AI involvement), can significantly influence trust independently when the information source is identical~\cite{Reis_Moritz}.
\major{In AI-powered tools, labeling plays a critical role, especially as users increasingly struggle to distinguish between human- and AI-generated content~\cite{llm_text_quality_1}.}
\major{In LLM-powered systems, the actual content source and the disclosed label can diverge, for example, AI-generated content may be labeled as human-authored.}
While prior research has independently examined the effects of information source (e.g., AI vs. human)~\cite{Walker, Johnson} and labeling~\cite{Reis_Moritz, Rae_Irene} on trust, there remains a critical gap in understanding how these two factors interact. Yet, both can significantly influence perceived trust in health information.
\major{This gap is especially important in high-stakes contexts like personal health, where trust directly influences individuals' health decision-making and behavior outcomes~\cite{journalmedia5020046}. 
Our work addresses this need by manipulating the content source and its disclosed label jointly to investigate their combined effects on people’s trust perception in health information, particularly in the era of LLMs.}

\begin{figure}
    \centering
    \includegraphics[width=\textwidth]{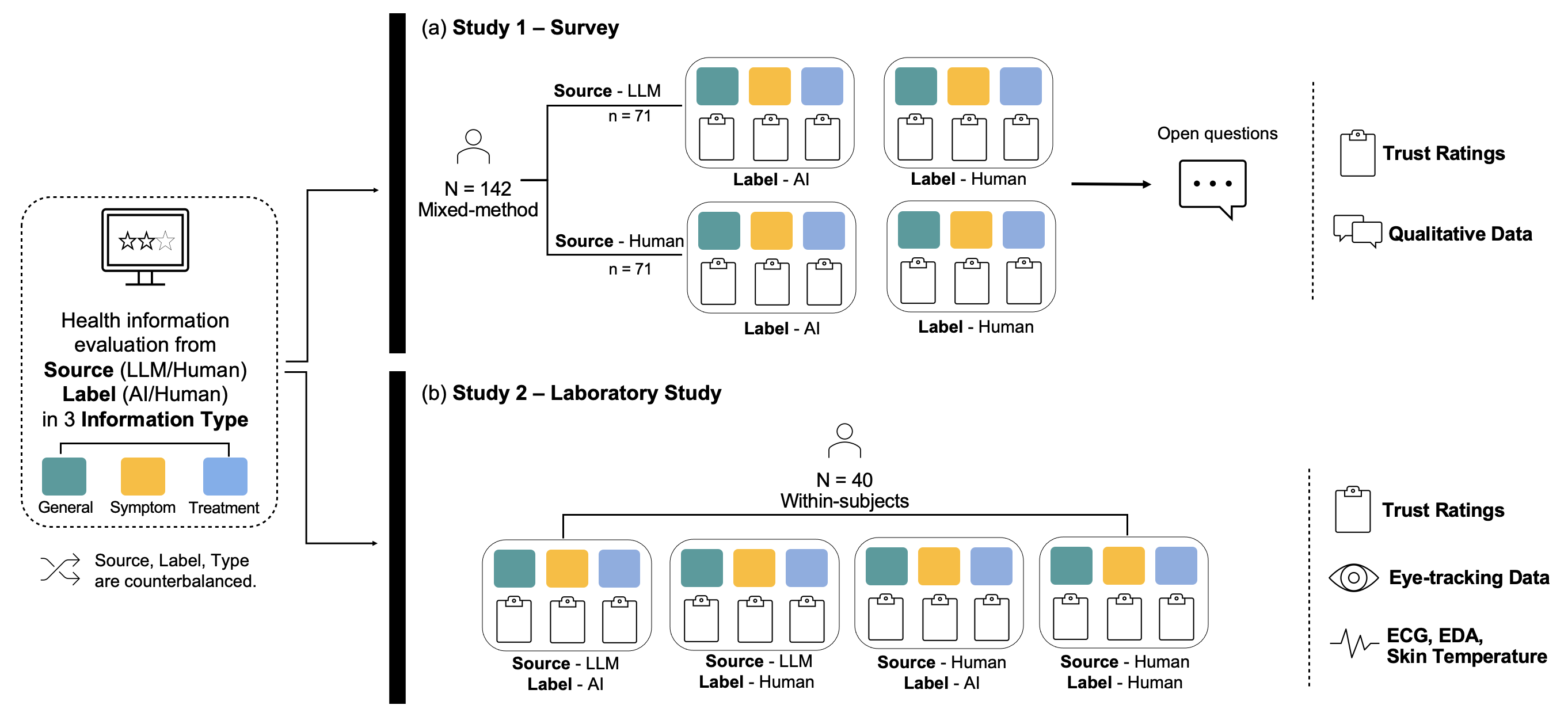}
    \vspace{-6mm}
    \caption{Visual summary of the studies in this paper. (a) Study 1: Mixed-methods crowdsourcing survey study to measure perceived trust; (b) Study 2: Within-subjects lab study to measure perceived trust, as well as behavioral and physiological responses.}
    \label{fig:procedure}
\end{figure}

To understand such joint effects of information sources and disclosed labels on people's perceived trust,
we ask: 
\textbf{(RQ1) How do the actual source, disclosed label, and type of personal health information influence people's perceived trust in online health information?}
To answer this research question, we employed a mixed-methods approach in Study 1 (see Fig~\ref{fig:procedure} (a)).
Specifically, we conducted an online crowdsourcing survey (N=142) using a 2x2x3 factorial design. 
Source (Human Professional vs. LLM) was treated as a between-subjects variable to minimize potential biases from participants directly comparing human and AI sources. 
In contrast, Label (Human Professional vs. AI) and health-information Type (General vs. Symptom- vs. Treatment-related) were within-subjects variables to enable a nuanced comparison of trust perceptions across different labeling and information types within the same participant. 
\major{This mixed design balanced the reduction of cross-condition biases with the sensitivity of within-subject comparisons. Participants rated their perceived trust in the health information they received using standardized self-report scales, which served as our primary trust measure outcome.}

Although self-reported measures \major{we adopted for Study 1} are widely used due to their simplicity and directness, 
Research by Chen et al.~\cite{Chen_Jing} and Kohn et al.~\cite{Kohn_Spencer} argue that self-reported trust measures are subjective, which makes them more vulnerable to biases like social desirability bias and the Initial Elevation phenomenon~\cite{Initial_Elevation_Phenomenon}.
These biases may compromise the reliability and validity of self-reported trust assessments.
With the growing use of sensing technologies and recent interest in Human–Computer Interaction research to draw on physiological sensing for designing or evaluating interactive systems~\cite{Chiossi2024},
\major{several prior studies~\cite{Ajenaghughrure, Akash_Kumar, Lim2022-gl} argue that behavioral and physiological data can provide a complementary perspective for understanding trust alongside self-reported measures.}
\major{Behavioral patterns such as eye movements and physiological responses, as assessed by Electrocardiogram (ECG)~\cite{Ajenaghughrure} and Electrodermal Activity (EDA)~\cite{eda_practice}, could reveal how individuals process information and make trust-related decisions in health contexts.}
For example, eye movement patterns, such as fixation duration and saccade behaviors, can indicate cognitive engagement with the information, while physiological responses like heart rate variability (HRV)~\cite{Ajenaghughrure, hrv_1, Ahmad_Muneeb} and skin conductance levels (SCL) can reveal emotional arousal and stress responses.
These \major{implicit} measures may further help interpret user trust perceptions~\cite{eda_practice, Ahmad_Muneeb}. 
Thus, exploring these behavioral and physiological indicators can contribute to a more comprehensive understanding of trust formation in digital health contexts~\cite{Akash_Kumar, Ajenaghughrure, gaze_eda_trust} and further, help develop strategies to enhance the trustworthiness of online health information, especially given the growing use of LLM-powered tools for health advice~\cite{t6,t7,t8}.

Building on RQ1, we adopt behavioral and physiological data as a complementary lens for understanding trust. 
We ask:
\textbf{(RQ2) Can behavioral and physiological signals be used to understand trust perceptions toward human and AI-generated health information?}
To address this research question, we conducted a laboratory study (Study 2, N = 40) using a 2 × 2 × 3 fully within-subjects design.
\major{We collected eye-tracking data (e.g., gaze patterns, pupil dilation) and physiological signals (e.g., ECG, EDA, and skin temperature) to examine whether these implicit signals vary as being manipulated by source and label.
Besides, we explored how these signals relate to participants’ self-reported trust perceptions.}
By allowing each participant to serve as their own control, this design minimized variability due to individual differences and maximized the robustness of condition-specific inferences.
\major{Importantly, participants were not informed that labels could be intentionally mismatched with the actual source (i.e., cross-labeled) in both studies.
This ensured that participants evaluated the health information and its disclosed label as presented, without being influenced by a heightened awareness of potential labeling errors,} thereby allowing us to more accurately assess their trust perceptions on both information itself and its labeling.

Online survey \major{(Study 1)} findings showed that the \major{(actual) source} of information significantly influenced trust perceptions, with participants displaying higher trust in LLM-generated health information compared with human professionals. 
Second, the labeling of the source played a crucial role: health information labeled as coming from human professionals led to significantly higher trust than information labeled as from AI, i.e., regardless of the actual source. 
\major{Third, the type of health question did not significantly affect trust, alone or in interaction with label and source. Together, these observations suggested that perceived trust is not influenced by the nature of the health query, and that the source and labeling of the health information are the main determinants.}
The \major{laboratory study \major{(Study 2)} supported the survey findings, with additional insights: gaze features, such as fixation, saccade, and pupil diameter,} varied significantly based on the source and labeling of health information.
Moreover, physiological features, such as heart rate variability \major{(HRV, measured as the root mean square of successive differences, RMSSD)} and skin temperature, differed when participants engaged with information with different labels. 
These findings indicated that the source and labeling of health information influence both behavioral and physiological responses. 
Further prediction tasks were performed based on behavioral and physiological data, yielding 0.35 $R^2$ for predicting trust scores and 73\% accuracy in classifying binary trust levels (high vs. low). Additionally, we achieved 65\% accuracy in classifying the source of health information. 
These results underscored the potential of leveraging behavioral and physiological signals \major{as complementary indicators} to understand trust perception toward human vs. AI-generated health information.

Our exploratory work offers two primary contributions: 
\textbf{(1)} We provided empirical evidence showing that trust in online health information is influenced both by its actual source and disclosed label. 
\textbf{(2)} We found that trust perceptions in personal health information vary at behavioral and physiological levels, offering \major{complementary insights beyond self-reported trust and helping to identify discrepancies between the explicit (i.e., self-reported) and implicit trust-related responses.}
\major{To our knowledge, this is one of the few studies that combine physiological (e.g., HRV, skin temperature) and behavioral (e.g., gaze) signals to understand trust in AI-generated health information.}
Our work highlights the importance of considering AI transparency labels when measuring trust in health information and the vulnerability of trust abuse to mislabeling. 
It further opens the possibility of verifying trust perceptions and inferring if and when to apply transparency labels based on sensed behavioral and physiological data.

\section{Related Work}

\subsection{Trust in Online Health Information Seeking}

\major{Trust is a multifaceted psychological construct essential to both interpersonal and human-technology interactions. 
Mayer, Davis, and Schoorman’s integrative ABI model of trust~\cite{organizational_trust_model} defines trust as a willingness to be vulnerable to the actions of another party, based on the expectation that the party possesses the performance (ability), intends to do good (benevolence), and adheres to a set of principles that the trustor finds acceptable (integrity).}
\major{Extending this concept to the digital age, Lee and See~\cite[p. 51]{trust_automation} define trust in technology as: ``An attitude that an agent will achieve an individual’s goal in a situation characterized by uncertainty and vulnerability''.}

In the context of health, trust is particularly important due to the sensitive nature of health information and its impact on health-related decision-making, which can have dire health consequences should it be incorrect~\cite{Ethical,journalmedia5020046}. 
Trust formation in health contexts is complex and influenced by both intrinsic and extrinsic factors, \major{including individual characteristics such as prior knowledge, health literacy, and external cues, such as source credibility, interface design, as highlighted by Vereschak et al.~\cite{Vereschak2024}.}
For instance, amounts of work~\cite{bates2006effect, Liu1, Sillence1, Singal, dutta2003trusted, Lucassen} indicated that the credibility of the information source is crucial, the design~\cite{8,9,14,21} and usability~\cite{24} of the health-related tools can significantly affect trust. 
User prior experience like familiarity levels~\cite{Sillence}, and user expectations~\cite{Guo2} influence trust perceptions as well. 
Moreover, users increasingly expect transparency, ethical AI practices, and data privacy, which further complicate trust calibration~\cite{ethics_1, Bansal, Sciascio, Haque}.

\major{To conceptually integrate the literature and these multi-level trust influences, we draw on the MATCH framework~\cite{responsible_ai}, a model that systematically captures the trustworthiness cues in human-AI communication.
Unlike trust models that focus either the trustee’s attributes (e.g., classic ABI model) or interface-level cues (e.g., MAIN model~\cite{Sundar2007TheMM}), the MATCH framework offers a more integrated account of how trust is formed in AI systems by collectively integrating content quality, interface design, and user heuristics.}

\major{Specifically, MATCH organizes trust into three components:
(1) Model Attributes. This dimension, drawn directly from the ABI model~\cite{organizational_trust_model}, refers to perceived ability and competence of the system. In our context, it relates to users' perceptions of the quality and reliability of the information itself.
It echoes the prior work revealing that the intrinsic quality of the information itself plays a critical role in shaping trust~\cite{2,4,9,14,t3}.
(2) Afforded Cues. These are extrinsic signals such as formatting, interface design or interaction patterns. 
Prior work shows that even subtle interface features like transparency labels~\cite{ai_transparency_trust,ai_people_heard_label} or content layout~\cite{ti,8,9,14,21} can significantly influence trust judgments.
(3) Trust Heuristics. MATCH uniquely accounts for the mental shortcuts users apply under uncertainty (e.g., quickly assessing that information labeled as ``human-generated'' is more trustworthy, or that ``AI-generated'' content is less reliable). 
It is often shaped by prior experience, health literacy, or cognitive and affective response collectively~\cite{trust_automation}.
In our work, we further interpret these heuristic processes through behavioral and physiological signals, such as gaze patterns that may reflect users’ implicit trust-related responses.
}

\major{Grounded in the MATCH model, this work examines how both source information taps into model attributes, how labeling functions as an afforded trustworthiness cue, and how behavioral and physiological signals reflect user cognitive heuristic and affective processing of health information toward the trust-related judgments. 
While existing studies have investigated trust in AI- vs. human-generated content, few have systematically decoupled actual source from disclosed label to assess their independent and combined effects. 
Our exploration builds on the MATCH model and extends prior work by isolating and manipulating both information source and labeling disclosure, allowing us to explore how these cues interact and shape trust formation in online health information seeking contexts.}



\subsection{Source and Label Transparency in the Age of LLMs}

The internet has become a vital resource for health information~\cite{Cline}, with websites like WebMD~\cite{WebMD} and Mayo Clinic~\cite{mayo} providing expert-curated content. 
The rise of LLMs like ChatGPT~\cite{chatgpt} has revolutionized access to online health information by offering conversational interactions to health queries~\cite{Dalton}.
Trust in these LLM-powered tools is influenced by various factors~\cite{Rheu}, including the perceived credibility of their responses, clarity of information, transparency about how the information is generated~\cite{transparent_ai}, and users' familiarity and experiences using such AI technologies~\cite{BICKMORE200521}.
Among these, information source (e.g., human-authored vs. AI-generated) plays a critical role in shaping trust. 
Research~\cite{trust_source_1, bates2006effect, Lucassen} has shown that trust is significantly affected by the perceived credibility of information source. 
While LLMs have been effective in providing health information~\cite{BICKMORE200521, ChatGPT_therapist}, concerns remain about their credibility and reliability. 
Although human professionals are traditionally viewed as authoritative and trustworthy due to their expertise~\cite{Kerstan,Broom2005}, studies like Logg et al.~\cite{Logg} showed that users may trust AI for specific tasks, and Shruthi et al.~\cite{shekar2024peopletrustaigeneratedmedical} indicated that people overtrust AI-generated medical responses. 
However, other research~\cite{Reis_Moritz, Kerstan} highlighted people's preferences for human-generated health advice, suggesting that trust varies based on context. 
Additionally, Montag et al.~\cite{trust_human_ai_eeg} found that trust in humans and AI may not be directly associated, suggesting people have distinct trust mechanisms for each. These varied trust levels underscore the complexity of trust formation towards information from human and AI sources.

Labeling of information sources plays an additional key factor in shaping trust perceptions in the era of LLMs. 
Jakesch et al.~\cite{ai_profile_text} demonstrated that \major{users perceive content as less trustworthy when it is labeled as AI-generated, even when the content quality is identical,} which indicates that labeling influences how users perceive trustworthiness. 
Similarly, Reis et al.~\cite{Reis_Moritz} found that perceived AI involvement significantly impacts trust in digital medical advice, \major{as participants in their study were less willing to follow health advice when they believed it was generated by AI rather than a human expert.}
Studies by Walker et al.~\cite{Walker} and Kerstan et al.\cite{Kerstan} have also shown that people tend to trust advice more when it comes from human professionals rather than from LLMs, especially when the source is explicitly stated. 
Yin et al.~\cite{ai_people_heard_label} found that while AI can create a sense of being heard, labeling content as AI-generated can reduce its perceived impact. 
These findings underscore how labels can significantly impact trust, even when AI performs tasks effectively. 
Furthermore, Scharowski et al. \cite{scharowski_certification_2023} explores the potential for AI certification labels (e.g., ``Digital Trust Label'' by the 2023 Swiss Digital Initiative), and finds that such labels can mitigate data-related concerns surfaced by end-users such as data protection and privacy, however this came at the cost of other concerns such as model performance, which poses it own challenges. 
Nevertheless, these works highlight that transparent communication about how AI systems operate and the data sources they use can further enhance or maintain trust among users~\cite{ai_transparency_trust, Logg}. 

As AI becomes more integral to health contexts, these work specifically explored the influence of source and labeling as critical extrinsic cues on trust in health information, offering insights for designing trustworthy LLM-powered health systems.
\major{Framed through the MATCH model~\cite{responsible_ai}, these effects reflect how users interpret afforded cues (e.g., disclosure labels) and model attributes (e.g., inferred competence or benevolence of a human vs. AI source) when assessing trust. Labels operate as interface-level factor that invoke trust heuristics, particularly under conditions of uncertainty. These trust dynamics highlight the importance of carefully designing how source and authorship are communicated in AI-powered health systems.}


\subsection{Behavioral and Physiological Signals for Understanding Trust Perception}

Traditional research on trust perception has heavily relied on self-reported assessments; however, many studies~\cite{Chen_Jing, Kohn_Spencer} suggest behavioral and physiological signals may add a relevant layer of information. Integrating these implicit measures helps offer \major{a complementary understanding of trust in} human and LLM-generated health information. 
\major{For example, research by Kenneth et al.~\cite{eyetracking_methods} shows that eye movement metrics like fixation, saccade, and pupil dilation provide insights into cognitive load and attention allocation during information processing. While these physiological indicators do not directly measure trust, they may reflect how users cognitively and affectively engage with content they perceive as more important, credible, or challenging.}
For instance, increased pupil dilation, linked to higher cognitive load~\cite{cognitive_load_physio} and emotional arousal, may suggest deeper cognitive processing, which may co-occur when individuals are evaluating information for trustworthiness or making health-related decisions. \major{Although the relationship between trust, cognitive, and affective responses is complex, monitoring these signals may help identify moments of increased scrutiny or hesitation, offering indirect cues about trust-related states.}
\major{As an example of such research, Ji et al.~\cite{physio_search, Ji_Kaixin} demonstrated that physiological signals, such as electrodermal activity, blood volume pulse, and gaze, vary meaningfully across different information processing activities (e.g., reading, speaking, listening) during information-seeking tasks.}
\major{Moreover, prior work has used behavioral data to explore how people engage with online news content, particularly in the context of misinformation. 
For instance, Abdrabou et al.~\cite{Abdrabou_Yasmeen} found that gaze and mouse movement patterns could help distinguish between user exposure to real versus fake news, achieving moderate accuracy in identifying subconscious engagement with misinformation. 
Similarly, Sumer et al.~\cite{SUMER2021106909} showed that eye-tracking data reflected differences in how users read and process true versus false news articles, suggesting that such behavioral signals can offer a comprehensive understanding of how people implicitly respond to varying degrees of information credibility.}
Studies~\cite{Lu_Yidu, gaze_eda_trust, Kohn_Spencer, eyetracking_methods, Sevcenko, Ayres2021-xe} demonstrate that distinct gaze patterns are linked to trust levels, with higher fixation counts and longer duration typically indicating focused attention, greater cognitive engagement, and trust in the information. 
\major{Saccades, characterized by the frequency and length of eye movements between fixations, often signal information verification processes~\cite{Lu_Yidu, Wang2001-hf, Wang_Lin}. 
These findings suggest that these multimodal implicit signals can be sensitive indicators of user cognitive effort and engagement, offering potential to infer user states such as trust or uncertainty in information processing contexts.
}

Physiological features such as ECG~\cite{Ajenaghughrure}, EDA~\cite{eda_practice}, and skin temperature~\cite{Ahmad_Muneeb} can be useful for understanding implicit responses related to trust. 
Heart Rate Variability (HRV), derived from ECG, reflects the level of stress and cognitive dissonance, with higher HRV indicating lower physiological arousal which is associated with relaxation, comfort, and higher trust levels~\cite{hrv_1,hrv_2,hrv_3}. 
EDA measures, including Skin Conductance Level (SCL) and Skin Conductance Response (SCR) are \major{similarly tied to} emotional arousal, where lower conductance is used to infer greater comfort and trust~\cite{eda_practice, gaze_eda_trust, Ahmad_Muneeb}. 
Similarly, changes in skin temperature \major{are thought to reflect engagement levels}, with higher temperature suggesting increased cognitive engagement with information~\cite{Ahmad_Muneeb}.
\major{As investigated by prior work~\cite{trust_automation}, trust perception, a complex, subjective cognitive and affective process,} can be assessed using models by analyzing physiological (e.g., ECG and EDA~\cite{Ajenaghughrure_modeling}, EEG~\cite{Akash_Kumar}) and behavioral (e.g., gaze patterns ~\cite{Lim2022-gl,Parikh2018EyeGF}) indicators. 
These models help reduce subjective bias and can provide real-time insights into trust responses, least of which is an additional verification means alongside self-reports. 

These behavioral and physiological signals provide insights into users’ implicit responses, capturing attention, emotional arousal, and cognitive engagement that may not surface in self-reports. 
\major{In our work, we explore whether implicit signals vary meaningfully across conditions of information source and labeling. We interpret these signals cautiously as indirect indicators that may correlate with trust.
Within the MATCH model~\cite{responsible_ai},} these sensing signals map onto trust heuristics component, reflecting how users internally process trustworthiness cues that influence trust.
Unlike explicit cues like source attributions, sensing signals help uncover how users process those cues implicitly, for example, when trust is assigned reflexively versus analytically. 
By revealing how trust is formed or challenged beneath explicit awareness, these signals complement extrinsic cues and help build a more comprehensive picture of trust in LLM-powered health contexts.


\subsection{Synthesis and Research Gap}

\major{As summarized in Table~\ref{tab:summary_prior_work}, prior research has largely treated source and label in isolation, and separately examined how information sources and disclosed labels influence trust in online information, but findings are mixed.
Some studies report that users trust human-generated content more due to perceived expertise and accountability~\cite{Kerstan, Walker}, while others show higher trust in AI-generated information, citing perceived consistency or objectivity~\cite{Logg, shekar2024peopletrustaigeneratedmedical}.
Research on labeling further shows that disclosing AI involvement often reduces trust even when content is identical~\cite{Reis_Moritz,ai_profile_text,ai_people_heard_label}.
However, few studies have systematically disentangled the effects of source and label together, or explored whether these effects vary across different information types in health contexts (e.g., general, symptoms, treatment).
}

\major{Moreover, prior research relies heavily on self-reported trust, which may not capture users’ implicit cognitive and emotional responses involved in trust judgment. 
Behavioral and physiological signals offer promising but underexplored means for revealing how users attend to, process, and evaluate health information beyond what they report, which can offer complementary insights into how trust is formed beyond self-reports.}

\major{
This leaves critical gaps (summarized in Table~\ref{tab:summary_prior_work}) in understanding how information source, labeling, and content type jointly influence both users' explicit trust (self-reports) and implicit responses (behavioral and physiological) in the context of LLM-generated health information.
To address this, our work draws on the MATCH framework~\cite{responsible_ai}, which integrates: 
Model Attributes (i.e., information source), 
Afforded Cues (i.e., disclosed label and information types), and 
Trust Heuristics (cognitive or emotional responses implicitly reflected in sensing signals).
This integrated approach allows us to investigate not only how trust varies across source, label, and information type, but also whether behavioral and physiological signals reflect trust-related judgments in implicit but meaningful ways when users engage with AI- and human-generated health information in LLM-powered contexts.
}

\vspace{-2mm}

\begin{table}[!ht]
\centering
\renewcommand{\arraystretch}{1.1}
\tiny
\setlength{\tabcolsep}{4pt} 
\begin{tabularx}{\textwidth}{@{}>{\raggedright\arraybackslash}m{1.8cm}|>{\raggedright\arraybackslash}m{3.7cm}|>{\raggedright\arraybackslash}m{4.2cm}|>{\raggedright\arraybackslash}m{2.8cm}@{}}
\toprule
\textbf{Manipulation} & 
\textbf{Source} & 
\textbf{Labeling} & 
\textbf{Source + Labeling} \\
\midrule
\multirow{6}{1.8cm}{\centering\textbf{Self-reports}} 
&  \textit{\textbf{Higher trust in AI than humans:}}\newline
\cite{Logg} (General context);\newline \cite{shekar2024peopletrustaigeneratedmedical} (Health context); 
& \textit{\textbf{Labels increase trust:}}\newline
\cite{scharowski_certification_2023} (General context);\newline
& \multirow{6}{3.0cm}{\centering\textbf{This work (Study 1):} Source+Label joint effects} \\
\cline{2-3}

& \textit{\textbf{Higher trust in humans than AI:}}\newline
\cite{Walker} (Binary decision-making);\newline
\cite{Kerstan,trust_source_1} (Health context);
& \textit{\textbf{AI-Human mixed labels decrease trust:}}\newline
\cite{ai_profile_text} (Marketing context);\newline
\textit{\textbf{AI labels decrease trust:}}\newline
\cite{ai_people_heard_label,Rae_Irene} (General context); \cite{Reis_Moritz} (Health context);
& \\

\midrule
\multirow{3}{1.8cm}{\centering\textbf{Self-reports + Sensing}}
& Trust differs by sources \textit{(Gaze Data in Fake News)}
& \multicolumn{2}{>{\centering\arraybackslash}m{6.6cm}}{
\multirow{3}{6.0cm}{
\centering\textbf{This work (Study 2):}\\Gaze+Physio in health context shows differences}} \\
\cline{2-2}

& Trust toward human and AI are not associated \textit{(EEG Data)} & \multicolumn{2}{c}{} \\
\bottomrule
\end{tabularx}

\vspace{-0mm}
\caption{Comparative synthesis of prior studies on source and label effects in trust perception and the research gaps we address in this work.} 
\label{tab:summary_prior_work}
\vspace{-3.6mm}
\end{table}

\section{Study 1: Online Survey}

\subsection{Study Methods}

\subsubsection{Design}
We conducted an online survey using a mixed 2 (IV1 - Actual Source: Human professionals vs. LLM) × 2 (IV2 - Disclosed Label: Human professionals vs. Artificial Intelligence) × 3 (IV3 - Information Type: General vs. Symptom vs. Treatment) factorial design to explore people's perceived trust in online health information. 
The source of health information (IV1) was set as a between-subjects variable to explore whether people have different trust perceptions based on the source (human professionals vs. LLM), which might inherently present information in distinct styles.
A within-subject design for the source could introduce biases in perceived quality and trustworthiness due to these stylistic differences. Additionally, using a between-subjects design for the source helps isolate the effect of labeling (IV2), making the findings clearer and more robust.
Conversely, for the label of the source (IV2) and the type of health information (IV3), we opted for a within-subjects design to allow direct comparisons of trust perception across different labels and types while keeping the source uniform for each participant. This approach reduces individual variability, ensuring a clearer separation of source effects on trust variances while enabling robust analysis of influences from labeling and types of health information.
Therefore, during the completion of the survey, each participant read the information either generated by human professionals or LLMs, and each of them experienced six distinct conditions.


\subsubsection{Health Information.}
\label{health_information}

Sets of health information (question and answer pairs) from human professionals were selected from an open-sourced dataset~\cite{MedQuAD} due to its diverse range of health questions, authored by certified professionals. This ensures the reliability and authenticity of the information used in this work.
\major{To produce comparable and consistent LLM-generated information, we used Generative Pre-trained Transformer 4 (GPT-4) model~\cite{gpt4}} (version: ``gpt-4-0125-preview'' through official API) and prompted it with selected health question and accompanying instructions (e.g., ``
Health question:
[question].
Please give answer to the above question within [wordcount] words?'') to generate answers of similar length to those from human professionals.
\major{To ensure consistency and mitigate potential misinformation, all LLM-generated responses were independently reviewed by two researchers using the corresponding human-authored answers as reference. The review criteria are consistency in length, format, topic relevance, and absence of harmful content. Only responses with full agreement were included, following established HCI practices~\cite{irr_scores}.}
The health information falls into three categories, \major{reflected in both the clinical process and the dataset’s validated taxonomy~\cite{MedQuAD}}:
\textbf{General information}:
provides answers to general health topics (e.g. ``Do you have information about weight control?'');
\textbf{Symptoms-related information}:
focuses on symptoms and potential diagnoses (e.g. ``What are the symptoms of burns?'');
\textbf{Treatment-related information}:
provides treatment options for specific conditions (e.g. ``What to do for burns?''). 
\major{
This categorization aligns with clinical practice, which commonly follows a three-stage diagnostic process~\cite{diagnostic_process_1,diagnostic_process_2}: assessment (general inquiry), diagnosis (symptom evaluation), and treatment planning (intervention). 
These types capture a progression from low- to high-stakes information, allowing us to explore whether trust perceptions vary by the nature of health content.}

Twenty-five questions were selected from each category resulting in a question set with 75 questions in total, ensuring a comprehensive representation of individual health questions.
The complete list of health information used in the study is included as \textbf{Supplementary Material}.


\subsubsection{Measures}
\label{measures}

\textbf{Demographics and prior experience.}
In the pre-survey, we collected participants' demographic information (age, gender, education, occupation) and their experience in online health information seeking, \major{using two questions: ``How often do you search for health information online?'' rated on a 5-point Likert scale from Never to Daily; and ``How long have you been using online sources for health information searching?'' with options ranging from Less than 1 year to More than 10 years.}

\textbf{Propensity of trust in technology (PPT)~\cite{ppt}} was used to assess inherent trust in technology before participants read the health information. It consists of 6 items looking at people's general trust in technology (e.g. ``I think it’s a good idea to rely on technology for help''). All items were scored on a 5-point Likert scale from 1 (Strongly Disagree) to 5 (Strongly Agree) (Cronbach’s \(\alpha = 0.71\)).

\textbf{eHealth and AI literacy.}
As part of the pre-survey, we also measured participants' literacy on eHealth and AI separately using two adapted questionnaires from eHEALS: The eHealth Literacy Scale~\cite{eHealth_literacy} and MAILS - Meta AI literacy scale~\cite{ai_literacy}. All the items were scored from 1 (Strongly Disagree) to 5 (Strongly Agree). The adapted measure for eHealth literacy has eight items with an example as ``I know where to find helpful health resources on the Internet'' (Cronbach’s \(\alpha = 0.88\)), and the adapted measure for AI literacy has ten items with an example item as ``I can distinguish if I interact with an AI or a real human'' (Cronbach’s \(\alpha = 0.76\)).

\textbf{Trust of online health information \cite{ti, Rowley2015StudentsTJ}} (\textbf{Trust Score})
During the formal study, participants completed the trust of online health information questionnaire to rate their trust level after reading each set of health information. It consists of 13 items (e.g. ``The information appears to be objective.''), each rated on a 5-point Likert scale from 1 (Strongly Disagree) to 5 (Strongly Agree) (Cronbach’s \(\alpha = 0.92\)). We aggregated and calculated the average value of all 13 items to obtain our perceived \textbf{Trust Score}. We use this score for further analysis throughout our work.

\textbf{Post-survey: three open-ended questions}
At the end of the survey, participants were asked to reflect on their trust perceptions through three open-ended questions. These questions explored their views on (a) general trust in LLM-generated information versus information from human professionals, (b) how they assess the credibility of online information, and (c) how the labeling of the health information source influences their perceived trust.

\begin{table}[!ht]
\centering
\renewcommand{\arraystretch}{0.99}
\scriptsize
\begin{tabularx}{\columnwidth}{p{4cm} >{\raggedright\arraybackslash}p{6cm} >{\raggedright\arraybackslash}p{1.5cm}}
\toprule
\textbf{Demographic} & \textbf{Categories} & \textbf{Numbers of Participants (\%)} \\
\midrule
Gender & & (N=142)
\\ 
 & Female & 83 (58.5\%)
\\ 
 & Male & 58 (40.8\%)
\\
 & Non-binary & 1 (0.7\%)
\\
\hline
Age & & 
\\
 & 18-24 & 91 (64.1\%)
\\
 & 25-34 & 38 (26.8\%)
\\
 & 35-44 & 9 (6.3\%)
\\
 & 45-54 & 2 (1.4\%)
\\
 & 65+ & 2 (1.4\%)
\\
\hline
Education & & 
\\
 & High school degree or equivalent & 24 (16.9\%)
\\
 & Bachelor’s degree & 67 (47.2\%)
\\
 & Master’s degree & 49 (34.5\%)
\\
 & Doctorate or higher & 2 (1.4\%)
\\
\hline
Professional Domain & & 
\\
 & Health and Medical Science & 17 (12.0\%)
\\
 & Science, Technology, Engineering, and Mathematics (STEM) & 35 (24.6\%)
\\
 & Business, Economics, and Law & 35 (24.6\%)
\\
 & Arts, Culture and Entertainment & 19 (13.4\%)
\\
 & Government and Public Sector & 3 (2.1\%)
\\
 & Education & 3 (2.1\%)
\\
& Other & 30 (21.1\%)
\\
\hline
Frequency of online health \\information seeking & & 
\\
 & Rarely & 27 (19.0\%)
\\
 & Sometimes & 77 (54.2\%)
\\
 & Often & 31 (21.8\%)
\\
 & Always & 7 (4.9\%)
\\
\hline
Duration of online health \\information seeking & & 
\\
 & Less than 1 year & 4 (2.8\%)
\\
 & 1-3 years & 24 (16.9\%)
\\
 & 3-5 years & 51 (35.9\%)
\\
 & 5-10 years & 45 (31.7\%)
\\
 & More than 10 years & 18 (12.7\%)
\\
\bottomrule
\end{tabularx}
\vspace{-1.6mm}
\caption{Characteristics of participants in the online survey.}
\vspace{-4.6mm}
\label{table:survey_demo}
\end{table}


\subsubsection{Participants}

Participants were recruited through the online crowd-sourcing platforms Prolific~\cite{Prolific} and institute recruitment systems. Our inclusion criteria included individuals over the age of 18 who are fluent in English, and they must have passed the attention check.
A power analysis conducted with G*Power 3.1~\cite{gpower} for a mixed-factor ANOVA design indicated that a minimum of 76 participants would be required to detect a small effect size (f=0.15), with an alpha level of 0.05 and \major{a power of 95\%.}

Table~\ref{table:survey_demo} shows a summary of participants' demographics.
142 participants (N=142) were recruited (F=83, M=58, NB=1), with 90.9\% falling in the 18-34 age bracket. 
Regarding educational backgrounds, 47.2\% had undergraduate degrees and 35.9\% held postgraduate qualifications. 
As for online health information-seeking experience, 26.7\% frequently used online sources, 54.2\% occasionally searched online, and 19.0\% rarely used online resources.


\subsubsection{Procedure}
\begin{figure}[!htbp]
\centering
\includegraphics[width=0.98\textwidth]{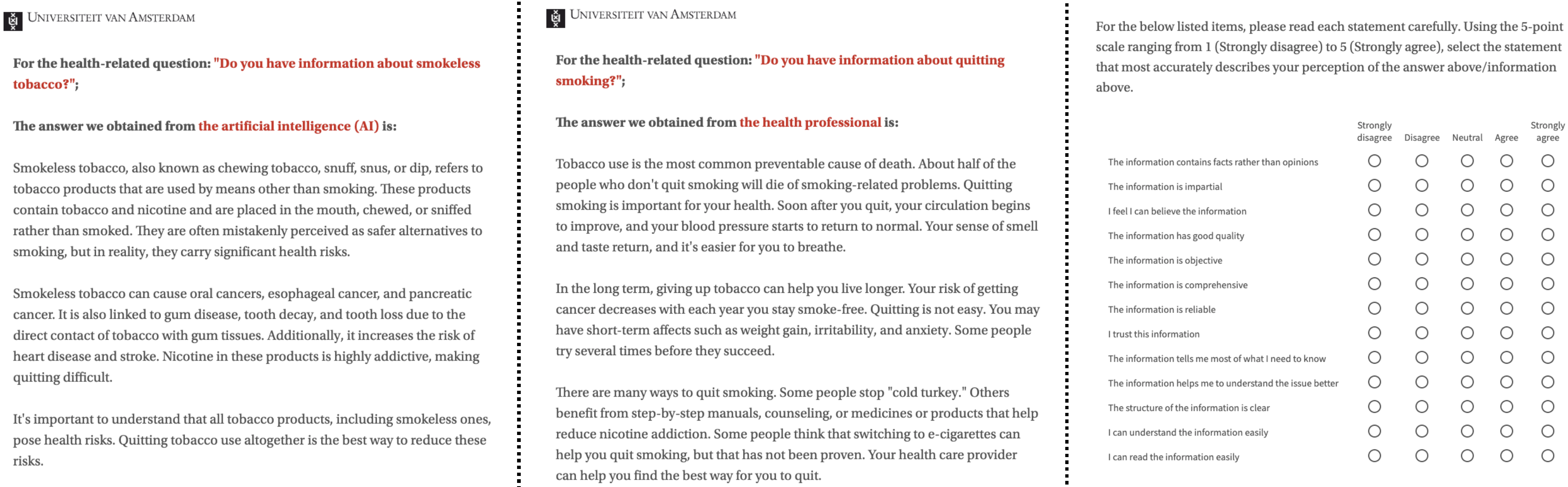}
\vspace{-0mm}
\caption{
Example reading task from the survey, showing a Q\&A pair with its assigned disclosed source label. 
Each participant read six Q\&A pairs: three labeled as from ``AI'' \textbf{(left)} and three labeled as from ``Human Professionals'' \textbf{(middle)}.
After each reading, participants rated their trust using the scale shown on the \textbf{right}. 
}
\label{fig:survey_example}
\end{figure}

The study design and procedure are outlined in Fig~\ref{fig:procedure}(a). 
Participants were first provided with detailed information about the study and gave informed consent in line with institutional guidelines. They provided demographic information and their experiences with online health information seeking. 
\major{A total of 75 health questions were used in the online survey, divided evenly into three categories: general health, symptom-related, and treatment-related (25 each) (Sec~\ref{health_information} ``Health information''). For each participant, six Q\&A pairs were shown: two randomly selected from each category. 
The survey study used a between-subjects design for the source of the information (AI- vs. human-generated) and a within-subjects design for the label (AI- vs. human-labeled). 
Both source and label orderings were counterbalanced based on a Latin square approach, ensuring that all condition combinations were evenly distributed across participants to mitigate order effects.}
\major{An illustrative example of the reading task interface during the survey is shown in Fig~\ref{fig:survey_example}.}
After reviewing each Q\&A pair, participants rated their perceived trust in the information.
At the end, participants completed a post-survey comprising three open-ended questions about their perceptions of the information source and its labeling.

Participation was voluntary and participants were monetarily compensated for a 30-minute session. 
\major{To ensure we avoid bots in our responses, we included an additional attention check where respondents needed to select a specific response to one question.}
Our study received approval from our institute's ethics and data protection committee.


\subsubsection{Data Analysis}
We conducted quantitative analyses to examine how the types of health questions, information sources, and labeling of sources influence trust perception in online health information. 
Initially, we confirmed the data's suitability for parametric tests by performing the Shapiro-Wilk test~\cite{SHAPIRO1965} for normality and Bartlett's test~\cite{Arsham2011} for homogeneity of variance; neither assumption was violated.
Next, we performed a mixed model, i.e., three-way mixed ANOVA~\cite{mixed_anova} to investigate differences in trust perceptions based on information sources, disclosed labels, and types of information. Since only one ANOVA was conducted, no correction for multiple tests was applied.
Following, post-hoc pairwise comparisons were conducted using t-tests with False Discovery Rate (FDR) correction~\cite{fdr_bh} to examine differences in trust between each pair of label and source combinations. 
To explore the relationship across variables, we also conducted Pearson correlation analyses~\cite{pearson} on two subsets of the data: one with human-sourced information and the other with LLM-sourced information (between-subjects independent variable). 
\major{Before analysis, we confirmed that the assumptions for Pearson correlation: normality, linearity, and absence of extreme outliers, were satisfied in the aggregated data.}
Bonferroni correction\cite{bonf_corr_1} was applied to account for multiple comparisons in both correlation analyses.

We conducted an inductive content analysis~\cite{elo2008qualitative} on the responses to three open-ended questions, focusing on identifying underlying themes that explain trust rather than counting frequencies. 
In the first stage, the first two authors created an initial set of codes using the qualitative analysis software ATLAS.ti~\cite{atlas_ti}. This initial codebook looked at respondents' varying perceived trust in AI and human professionals, their reasons for trusting or distrusting, and how they typically evaluate the credibility and trustworthiness of information. Following this, both coders independently open-coded the responses, remaining open to new observations and emerging codes. Similar codes were merged, unclear ones were refined, and earlier responses were re-coded as needed. 
As the analysis progressed, recurring factors emerged across different questions, allowing us to develop common themes that spanned all three sets of responses.

\subsection{Quantitative Findings}


\subsubsection{Descriptive Statistics}

As shown in Table~\ref{tab:survey_descriptive}, participants demonstrated a positive propensity to trust in technology, with an average score of 3.85 (SD=.72), indicating a positive attitude towards technology.
The average eHealth literacy score was 3.62 (SD=.87), indicating that participants are relatively capable of using online health resources. 
AI literacy was also high, with an average score of 3.81 (SD=.92), reflecting a favorable understanding of AI technology. 

\begin{table}[!ht]
\centering
\renewcommand{\arraystretch}{1.2}
\scriptsize
\setlength{\tabcolsep}{3pt} 
\begin{tabular*}{\textwidth}{@{\extracolsep{\fill}}p{2cm}p{6cm}p{2cm}p{2cm}@{}}
\hline
\textbf{} & \textbf{Measures} & \textbf{Mean} & \textbf{SD} \\ \hline

\multirow{3}{*}{Pre-survey} 
 & Propensity of trust in AI technology (PPT) & 3.85 / 5 & .72 \\ \cline{2-4} 
 & eHealth literacy & 3.62 / 5 & .87 \\ \cline{2-4}  
 & AI literacy & 3.81 / 5 & .92 \\

\hline

\textbf{} & \textbf{Conditions} & \textbf{Mean} & \textbf{SD} \\ \hline

\multirow{8}{*}{Trust score} 
 & Source (Human) \& Label (Human) & 4.01 / 5 & .45 \\ \cline{2-4} 
 & Source (Human) \& Label (AI) & 3.76 / 5 & .49 \\ \cline{2-4}
 & Source (LLM) \& Label (Human) & 4.07 / 5 & .47 \\ \cline{2-4} 
 & Source (LLM) \& Label (AI) & 3.87 / 5 & .44 \\ \cline{2-4} 
 & Source (Human), regardless of Label & 3.89 / 5 & .84 \\ \cline{2-4} 
 & Source (LLM), regardless of Label & 3.97 / 5 & .81 \\ \cline{2-4} 
 & Label (Human), regardless of Source & 4.04 / 5 & .46 \\ \cline{2-4} 
 & Label (AI), regardless of Source & 3.82 / 5 & .47 \\ 
\bottomrule 
\end{tabular*}
\caption{Descriptive statistics of the online survey.}
\label{tab:survey_descriptive}
\vspace{-2.0mm}
\end{table}

In terms of trust perception, the trust scores (based on the aggregate Trust Score described in Sec~\ref{measures}) varied depending on the source and label of the information. For information both sourced from and labeled as human, the average trust score was 4.01 (SD=45). When the information was sourced from humans but labeled as AI, the trust score decreased significantly to 3.76 (SD=.64). In contrast, information sourced from LLM but labeled as human received the highest trust score of 4.07 (SD=.47), while information sourced from AI and labeled as LLM had a trust score of 3.87 (SD =.44). 
These findings highlight the ways in which both the source and labeling of information can impact trust perceptions, with a clear indication that labeling of the sources plays a role in shaping trust, potentially even more than the actual source of the information.

\begin{figure}[!htbp]
\raggedright 
\centering
\includegraphics[width=0.999\textwidth]{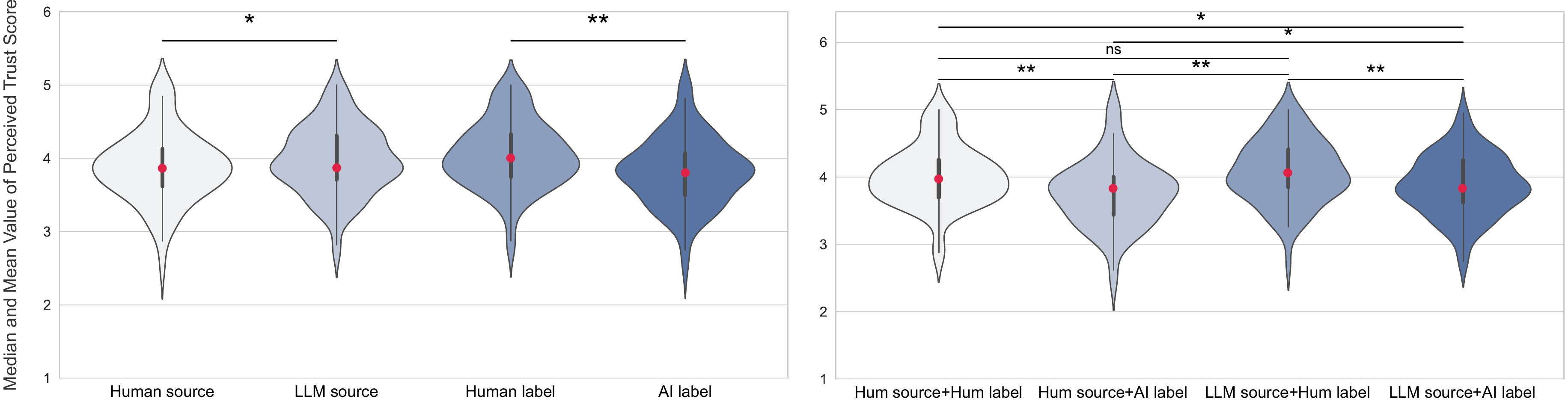}
\caption{\textbf{Left:} Perceived trust score in information by sources regardless of labels, and by labels regardless of source from the three-way mixed ANOVA without correction.
\textbf{Right:} Post hoc pairwise comparisons on perceived trust score based on different source and label conditions using t-test with False Discovery Rate (FDR) correction. 
\major{
Each plot shows the score density (width), with the red dot indicating the mean, the black line as the median, and thick bars representing the interquartile range (IQR). Horizontal lines indicate significance (**$p$<.01, *$p$<.05, ``ns'': no significance).
}}
\label{fig:survey_analysis}
\end{figure}

Our mixed model analysis compared differences in trust levels among the source, label, and health information types. Findings are shown in Table~\ref{tab:survey-trust_results} and Fig~\ref{fig:survey_analysis}, and together highlight how people perceive and trust health information manipulated by sources and labels.


\begin{table}[!h]
\centering
\renewcommand{\arraystretch}{1.2}
\scriptsize
\setlength{\tabcolsep}{2pt} 
\begin{tabular*}{\textwidth}{@{\extracolsep{\fill}}p{1.8cm}p{3.8cm}p{1.6cm}p{1.4cm}p{2.4cm}p{1.0cm}@{}}
\hline
\textbf{Outcomes} & \textbf{Conditions} & \textbf{Statistics} & \textbf{P-value} & \textbf{Effect size} & \textbf{Sig} \\ \hline
\multirow{3}{*}{Trust score} 
 & Source (Human vs. LLM) & 2.27 & .024 & .14 (medium) & * \\ \cline{2-6} 
 & Label (Human vs. AI) & -6.50 & .000 & -.39 (medium)  & ** \\ \cline{2-6} 
 & Type of health information & 0.67 & .505 & .05 (small) &  
 \\ 
\bottomrule
\end{tabular*}
\caption{Results from the three-way mixed ANOVA analysis on the trust score without data correction. (**$p$<.01, *$p$<.05)}
\label{tab:survey-trust_results}
\vspace{-5.0mm}
\end{table}

\subsubsection{Participants Gave Higher Trust to Health Information Sourced from LLM than from Human Professionals.}

The impact of the information source (human professionals vs. LLM) on trust in health information was analyzed by a three-way mixed ANOVA. The results showed significant differences in trust levels between sources: statistics=2.27, p=.024, effect size=.14.
This suggests that information sources significantly influence overall trust in health information.
Specifically, participants reported trusting information from LLM more than human professionals, with an average trust score for LLM-sourced information of 3.97 (SD=.81), compared to 3.89 (SD=.84) for information from human professionals.
\major{
Although perceived trust does not imply factual accuracy, our findings reflect a growing acceptance of AI-generated health content and shifting attitudes toward it relative to advice from human professionals.
}

\subsubsection{Participants Gave Higher Trust Ratings to Health Information Labeled as from Human Professionals compared to Labeled as from AI.}
Except for the factor of ``source'', the labeling of information sources influenced trust perception significantly.
Participants perceived significantly lower trust in health information labeled as from AI compared to that labeled as from human professionals, as indicated by a mixed model ANOVA (statistics=-6.50, p<.001, effect size=-.39), with an average trust score for information labeled as from human professionals of 4.04 (SD=.46) and 3.82 (SD=.47) for information labeled as from AI.
We also observed no significant difference in trust between human-labeled information from human sources (M=4.01, SD=.45) and LLM sources (M=4.07, SD=.47).
These results suggested that while LLM-generated information is generally trusted, the perceived trust still leans in favor of human-associated information when directly compared.


\subsubsection{The Type of Health Information \major{Does Not} Affect Participants' Trust Perception in Information.}
Additionally, we explored how trust varied across different categories of health information. There was no significant effect found (statistics=0.67, p=.505, effect size=.05). This suggests that the type of health question does not influence people's trust levels in health information.
The interaction effect between the label of the information source and category of information was not significant as well (statistics=-.51, p=.613, effect size=-.15).
This implies that the influence of labeling on trust does not vary across different types of health information.


\subsubsection{Correlation Analysis}
\begin{figure}[!htbp]
\centering
\includegraphics[width=0.98\textwidth]{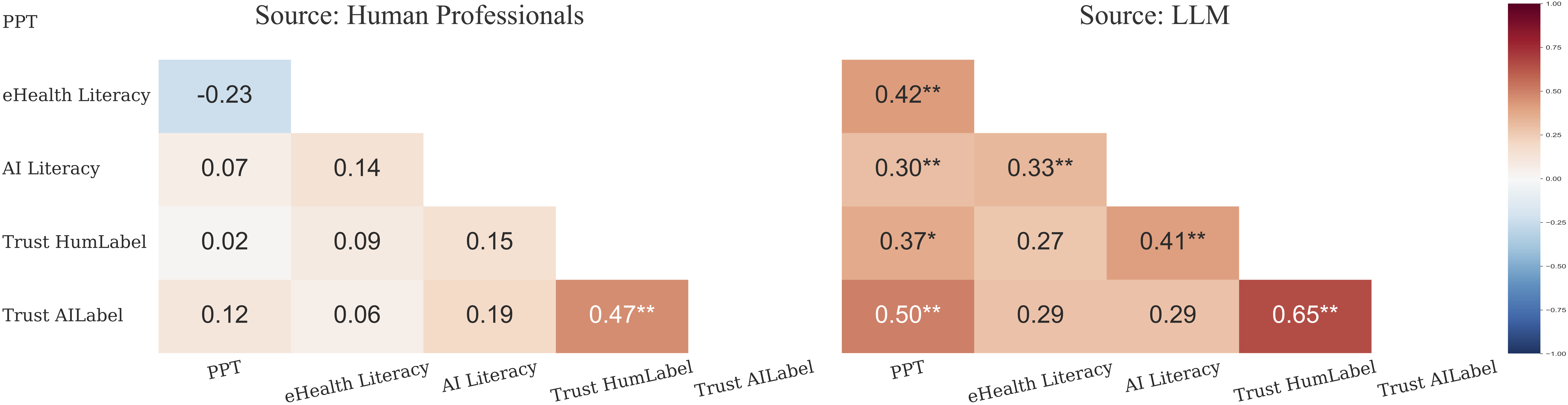}
\vspace{-0mm}
\caption{Pearson correlation with Bonferroni correction among the key variables in the online survey.(**$p$<.01, *$p$<.05) 
Note: 
``HumLabel'': information with human label regardless of the actual source.
``AILabel'': information with AI label regardless of the actual source.}
\label{fig:correlation}
\end{figure}

Given that the mixed ANOVA indicated no significant effect of the type of health information on the trust perceptions, the repeated measures were averaged into a single observation for each participant. This simplification allowed us to conduct a Pearson correlation analysis~\cite{pearson} to examine the general relationships between key variables in the online survey. The results, illustrated in Fig~\ref{fig:correlation}, revealed distinct patterns of trust in health information from different sources.
For information sourced from human professionals, trust in human-labeled information showed a moderate positive correlation with trust in AI-labeled information ($r(142)=0.47, p<0.01$). However, other relationships, such as those involving eHealth literacy and AI literacy, exhibited weak or negligible correlations.
In contrast, for information sourced from LLMs, we observed stronger correlations across multiple variables.
Trust in human-labeled information showed a strong positive correlation with trust in AI-labeled information (r(142)=0.65,p<0.01), AI literacy (r(142)=0.41,p<0.01), and the propensity of trust in AI ($r(142)=0.37,p<0.05$).
Additionally, the propensity of trust in AI correlated to trust in AI-labeled information (r(142)=0.50,p<0.01), eHealth literacy ($r(142)=0.42,p<0.01$), and AI literacy ($r(142)=0.30,p<0.01$) 
AI literacy positively correlated with eHealth literacy ($r(142)=0.33,p<0.01$).
These results highlighted a consistent influence of labeling on participants' trust across different sources.


\subsection{Qualitative Findings}
\label{sec:qual}

We received a total of 426 free-text responses (142 for each question). 
In this section, we present our findings with four themes. We found that participants' trust in AI versus humans is shaped by their inherent trust predispositions (Section~\ref{qualititive: inherent trust}) and their perceived source of knowledge for each agent (Section~\ref{qualititive: source}). 
Additionally, participants value human consciousness as a factor contributing to greater trust (Section~\ref{qualititive: consciousness}), and the presentation of information also influences their trust (Section~\ref{qualititive: presentation}).

\subsubsection{Predisposition toward AI and Humans Influences Trust}
\label{qualititive: inherent trust}
Survey respondents demonstrated a predisposition to trust either AI or humans, independent of the content or source of the information. However, there were individual differences in this inclination. Some respondents were optimistic about AI technology, regularly using and trusting AI in their daily lives. 
They perceived no difference in reliability between AI and human professionals, and some even trusted AI more.
Conversely, some respondents expressed significant reservations about AI, doubting its readiness to address serious topics, especially in sensitive fields like healthcare.
One respondent noted, \textit{``I don't trust AI, and the quick push in its advancements is dangerous; at the very least, it should be limited in specific fields such as health.''} 
Privacy concerns and the risks of AI-driven health advice reinforced such skepticism, leading to more critical evaluation of AI recommendations.
This underlying predisposition toward AI or human professionals also shaped respondents' views on labeling. 
Some participants expressed a preference for human-labeled content, with one stating, \textit{``AI label makes me trust it less and view the information more critically than if it came from a human professional.''} 
However, not all respondents allowed their predispositions to dictate their trust. Others placed less emphasis on labels, focusing instead on verifying information from multiple perspectives rather than relying solely on the source. As one respondent explained,
\textit{``The label doesn't affect how I interact with it, and my trust wouldn't be based solely on the label.''}

\subsubsection{Perceived Source of Knowledge Influences Trust}
\label{qualititive: source}
Survey respondents' trust in AI or human professionals was shaped by their perceptions of where each derives its knowledge.
One respondent explained,  \textit{``I would trust a human professional more, since he has learned factual information in school. An AI has learned from multiple sources online, not only factual ones, so that is why I would trust them a bit less.''}. In contrast, some respondents believed that AI can learn from \textit{``more databases and the most important points that all research brought up''}, potentially making it more knowledgeable than a single human expert. 
These differing views on the origins of human and AI’s knowledge contributed to varying levels of trust.
Some respondents took a more balanced stance, recognizing that both AI and human professionals are susceptible to biases and errors.
As one respondent commented, \textit{``While information from a human professional may need correction due to incomplete knowledge, information from AI might contain errors due to gaps in its training data.''} 
Consequently, many respondents shared that they would evaluate both sources of information with equal care, relying on their own experiences to evaluate the content's credibility. Additionally, some respondents expressed a preference for combining information sources, such as cross-checking information or using AI as a complementary tool to support human decision-making.


\subsubsection{The Human Touch Builds Greater Trust than AI}
\label{qualititive: consciousness}
Survey respondents highlighted that, due to the absence of consciousness and empathy in AI, they trusted human professionals more, particularly in healthcare contexts.
Many respondents emphasized that AI lacks the ability to evaluate information with awareness.
As one respondent commented, \textit{``Unlike human, AI doesn't know the difference between good or bad quality.''} In contrast, many respondents emphasized that human professionals have \textit{``years of medical education and experience with real-life cases''} to inform their decisions, something that AI cannot replace despite its access to vast information.
This absence of consciousness made respondents very skeptical about AI's capability to offer reliable health advice. 
The issue extended beyond decision-making to interpersonal interactions.
Respondents valued the sense of responsibility and ethical obligation that human professionals carry, with one noting,
\textit{``I trust the information from the human professional more because they are human and have moral and professional obligations about not giving misinformation.''}
Additionally, human-to-human interaction offered a sense of personalized care, making respondents feel their symptoms are better understood. In contrast, AI lacked this human touch, and its absence of empathy and accountability led respondents to trust it less.

\subsubsection{Presentation of Information Influences Trust}
\label{qualititive: presentation}
Information presentation was highlighted as an advantage of AI, which increased respondents' trust. They mentioned that when evaluating health information, factors such as the design of the user interface, the length of the information, the visible publication date, and the clarity of language were important. 
Compared to human professionals, AI was often perceived as providing simpler, more structured, and user-friendly information. Respondents appreciated that AI’s answers were clearly explained and easy to understand.
Additionally, the objective tone of AI responses further boosted respondents' trust. These elements collectively enhanced AI's explainability. As one respondent noted, \textit{``When I receive information from a human professional, I expect it to contain more academic language, which is harder to understand and less explanatory. Information from AI, however, uses simpler words and is easier to understand.''}


\section{Study 2: Laboratory Study}

\major{Study 1 demonstrated that the factors of actual source and disclosed label both affect people's perceived trust (self-reported) in health information.}
To further understand the process and user behaviors involved in forming trust perceptions, we conducted an in-person experiment. This study explored how health information from different sources and labels affects people’s behavioral and physiological states.

\subsection{Study Methods}


\subsubsection{Design}

Similarly to the online survey study, we utilized a within-subjects 2 (IV1 - Information Source: Human Professional vs. LLM) x 2 (IV2 - Disclosed Label: Human Professional vs. Artificial Intelligence) x 3 (IV3 - Information Type: General vs. Symptom vs. Treatment) factorial design tested in a controlled, laboratory environment (as shown in Fig~\ref{fig:procedure}(b)).
Different from Study 1, participants experienced all 12 distinct conditions for this in-person experiment, enabling direct comparisons between human and LLM-generated health information.
\revise{We opted for a within-subject design for all independent variables to facilitate a nuanced analysis of participants’ behavioral and physiological responses across different conditions. Specifically, for the source of information (IV1), we aimed to observe whether participants exhibited different behavioral (e.g., gaze patterns) and physiological (e.g., heart rate, skin conductance) signals when reading information attributed to human versus LLM sources. 
While these sources may differ in presentation styles, it is also possible that participants' trust perceptions were influenced more by their belief about the source of the text (human vs. AI) rather than the actual content or style. 
A within-subject design was critical for disentangling these effects, as it allowed each participant to serve as their own control, reducing variability across conditions and enabling a clearer examination of these factors.}
Participants rated their perceived level of trust for each set of health information while their eye-tracking data (gaze positions and pupil diameter) and physiological responses (ECG: Beats Per Minute (BPM), Beat-to-Beat Interval (BPI), Root Mean Square of Successive Differences (RMSSD); EDA: Skin Conductance Level (SCL) and Response (SCR); Skin Temperature) were recorded throughout the tasks.

To address our second research question, we explore whether behavioral and physiological signals can be used as \major{complementary indicators} to understand trust perceptions toward human and AI-generated personal health information. In addition, we set up two prediction tasks that make use of the sensed data: (1) predicting participants' trust in health information by both regression on perceived trust scores and binary classification on trust level (high vs. low); and (2) classifying the actual source of the health information.


\begin{figure}[!htbp]
\centering
\begin{minipage}[b]{0.999\textwidth}
    \centering
    \includegraphics[width=\textwidth]{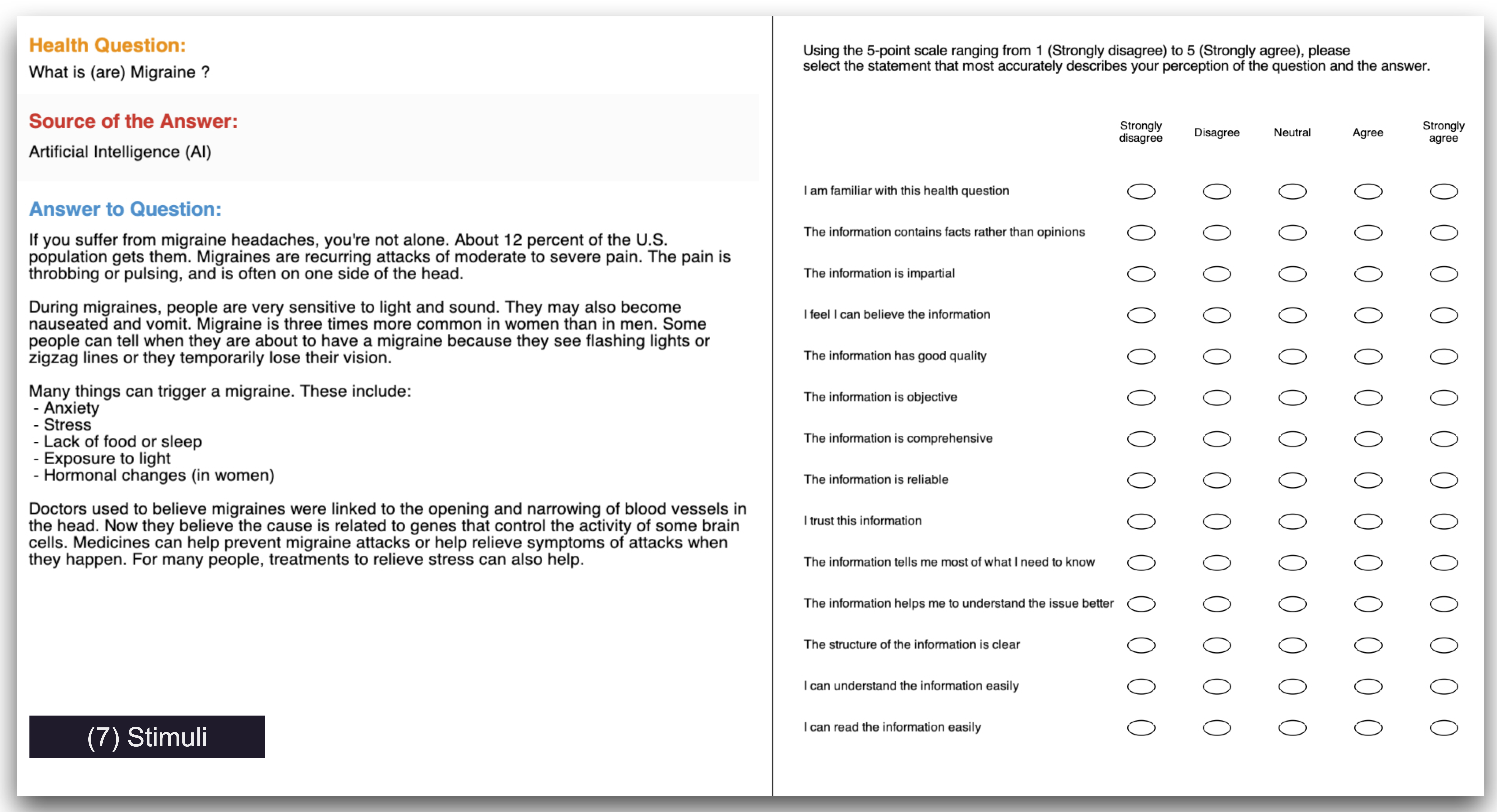}
    \label{fig:stimuli}
    \vspace{-2mm}
\end{minipage}
\begin{minipage}[b]{0.999\textwidth}
    \centering
    \includegraphics[width=\textwidth]{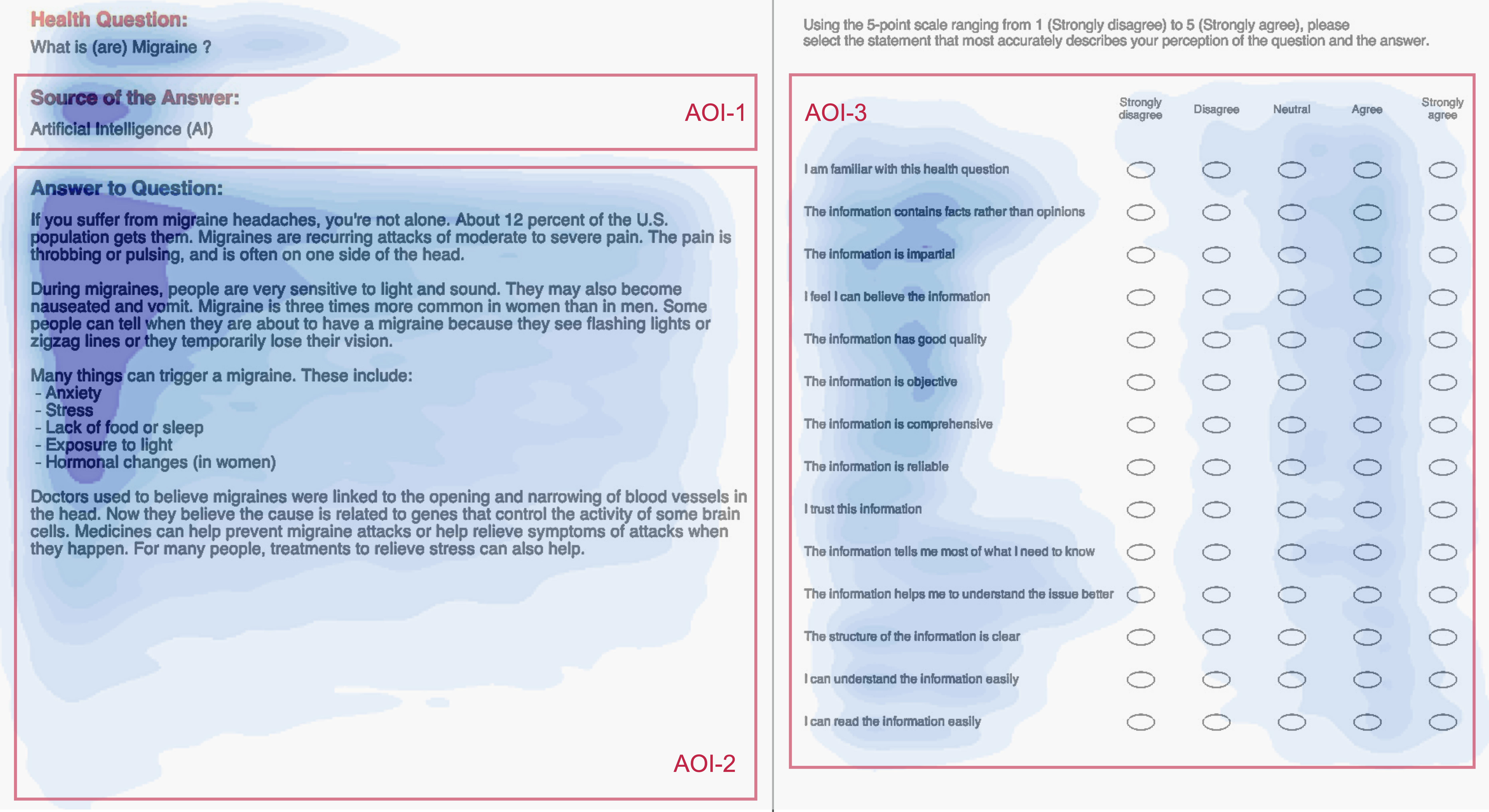}
    \label{fig:gaze_heatmap}
\end{minipage}
\vspace{-4mm}
\caption{\textbf{Top:} An example of text stimulus displayed on the monitor. \textbf{Bottom:} Heatmap of the gaze points on stimuli. Three AOIs are predefined: AOI-1 is the area for presenting disclosed label; AOI-2 is the area for presenting health information; AOI-3 is the area to rate the perceived trust in health information.}
\label{fig:stimuli-lab-study}
\vspace{-2mm}
\end{figure}


\subsubsection{Stimuli and Apparatus}
We developed a web interface that displays the health information (question and answer pair) and the questionnaires for participants to rate their trust scores (see Fig~\ref{fig:stimuli-lab-study}). The health information was identical to the material used in Study 1, as described in Section~\ref{health_information}. Each set of health information was labeled as being generated either by ``Human Professionals'' or ``Artificial Intelligence'', regardless of the actual source.

We used a PHILIPS (full HD, 1920*1080, 100 Hz) monitor to display the stimuli.
The eye movements and pupil diameter (PD) data were recorded by Tobii Pro Fusion eye tracker. The remote eye tracker was attached to the bottom of the monitor and connected to a computer (Windows, core Intel i5, 16GB RAM) running the Tobii Pro Lab software~\cite{TobiiProLab}.

Physiological signals, including ECG, EDA, and skin temperature, were measured using a BioSemi amplifier~\cite{physio_devices} (as shown in Fig~\ref{fig:teaser}). 
ECG was captured through a disposable 3M Red Dot in LEAD-II configuration, EDA was measured with electrodes attached to fingers, and skin temperature was monitored with a miniature Pt1000 sensor, all at a 24-bit resolution and 1000 S/s sampling rate. 
These data were collected using software FysioRecorder version 2.1~\cite{physio_devices}.
Data recording was initiated through a central recording application developed in PsychoPy~\cite{psychopy}, connecting to sensors via IP addresses to simultaneously capture synchronized ECG, EDA, skin temperature, and eye-tracking signals.


\subsubsection{Self-reported Measures}
\label{measure_lab}

We collected several self-reported measures, consistent with those used in Study 1 described in Sec~\ref{measures}. 
These included demographics, prior experience with online health information and AI, the propensity to trust technology (PPT), eHealth, and AI literacy.
Additionally, we assessed the perceived reliability of AI and human professionals using a single item for each: ``How reliable do you find AI/Human Professionals?'' Responses were captured on a 5-point Likert scale, ranging from 1 (Not at all) to 5 (Extremely). They were collected before the formal reading task.

During the reading task, we repeatedly measured the participants' 1) familiarity level with each given health question and 2) their perceived trust score in health information~\cite{ti, Rowley2015StudentsTJ}, after they completed each stimulus.


\subsubsection{Machine Learning: Setup and Approach}
\label{ml_analysis}

We performed binary classification to predict information sources and applied both regression and classification \revise{(i.e., binary and three-class classification)} for \revise{trust scores}. The perceived trust score (see Sec~\ref{measure_lab}), as an aggregate numerical rating based on the ``trust of online health information questionnaire''~\cite{ti, Rowley2015StudentsTJ}, naturally lends itself to regression. However, this approach can be challenging to interpret given that trust is an aggregate and overall complex construct. On the other hand, trust classification simplifies interpretation but introduces an arbitrary split between high and low trust levels. \revise{To address this, we pre-processed the original trust scores into high and low categories using the median value as a threshold for binary classification. For the three-class classification, we divided the trust scores into low, medium, and high categories based on tertiles, creating balanced splits that account for the distribution of scores.}

We used several common machine learning algorithms as suggested in prior research~\cite{Ajenaghughrure_modeling}, including single models (i.e., logistic regression (LR), random forest (RF), support vector machines (SVM), multi-layer perceptron (MLP), linear regression, ridge regression, random forest-based regression), and ensemble methods (i.e., boosting, voting, stacking and bagging). Models were built using hand-crafted gaze features (i.e., fixations, saccades, pupil diameter) and physiological signals (i.e., BPM, BPI, RMSSD, SCL, SCR, and skin temperature).

We experimented with three feature sets: Gaze-only, Physiology-only, and Gaze + Physiology. These sets trained and evaluated the selected models to determine how effectively they could predict participants' perceived \revise{trust scores} and classify the source of information.
We set the ``random state''~\cite{random_state} parameter to ensure result consistency and used the ``grid search''~\cite{grid_search} technique to find the optimal hyperparameters of the models. 
\major{We only consider user-independent models to ensure that any predictions generalize across all participants, despite well-known challenges in generalizing using peripheral physiological features~\cite{Alamudin2012}.}
\major{To achieve this, we adopted a Leave-One-Subject-Out (LOSO) cross-validation approach~\cite{LOSO}, where in each iteration, one participant’s data was held out for testing, and the remaining data was split 80/20 for training and validation. This setup ensures robust user-independent models.}
The performance of the regression models (for trust score prediction) was evaluated by Mean Squared Error (MSE) and Coefficient of Determination ($R^2$).
The performance of the classification models (for trust level and information source) was assessed with the Macro-F1~\cite{marco_f1} score as the average of the validations.


\begin{table}[!ht]
\centering
\renewcommand{\arraystretch}{0.96}
\scriptsize
\begin{tabularx}{\columnwidth}{p{5cm} >{\raggedright\arraybackslash}p{6cm} >{\raggedright\arraybackslash}X}
\toprule
\textbf{Demographic} & \textbf{Categories} & \textbf{Numbers of Participants (\%)} \\
\midrule
Gender & & (N=40)
\\ 
 & Female & 23 (57.5\%)
\\ 
 & Male & 16 (40.0\%)
\\
 & Non-binary & 1 (2.5\%)
\\
\hline
Age & & 
\\
 & 18-24 & 23 (57.5\%)
\\
 & 25-34 & 14 (35.0\%)
\\
 & 35-44 & 1 (2.5\%)
\\
 & 45-54 & 1 (2.5\%)
\\
 & 65+ & 1 (2.5\%)
\\
\hline
Education & & 
\\
 & Bachelor’s degree & 18 (45.0\%)
\\
 & Master’s degree & 17 (42.5\%)
\\
 & Doctorate or higher & 5 (12.5\%)
\\
\hline
Professional Domain & & 
\\
 & Health and Medical Science & 2 (5.0\%)
\\
 & Science, Technology, Engineering, and Mathematics (STEM) & 11 (27.5\%)
\\
 & Business, Economics, and Law & 8 (20.0\%)
\\
 & Communication, Arts, Culture, and Entertainment & 7 (17.5\%)
\\
 & Education and Social Science & 7 (17.5\%)
\\
& Other & 5 (12.5\%)
\\
\hline
Frequency of online health \\information seeking & & 
\\
 & Rarely & 6 (15.0\%)
\\
 & Sometimes & 25 (62.5\%)
\\
 & Often & 7 (17.5\%)
\\
 & Always & 2 (5.0\%)
\\
\hline
Frequency of using AI tools & & 
\\
 & Never & 2 (5.0\%)
\\
 & Rarely & 5 (12.5\%)
\\
 & Sometimes & 9 (22.5\%)
\\
 & Often & 18 (45.0\%)
\\
 & Always & 6 (15.0\%)
\\
\hline
Duration of online health \\information seeking & & 
\\
 & Less than 1 year & 4 (2.8\%)
\\
 & 1-3 years & 24 (17.0\%)
\\
 & 3-5 years & 50 (35.5\%)
\\
 & 5-10 years & 45 (31.9\%)
\\
 & More than 10 years & 18 (12.8\%)
\\
\bottomrule
\end{tabularx}
\caption{Characteristics of participants in the lab study.}
\label{table:lab_demo}
\end{table}

\subsubsection{Participants}

For the in-person experiment, we used the same inclusion criteria as in Study 1 (age above 18 who are fluent in English).
Participants were recruited through the institute's recruitment system.
A power analysis using G*Power 3.1~\cite{gpower} for a within-factor ANOVA indicated that at least 28 participants were required to detect a medium effect size seen in Study 1 (f=0.24), with an alpha level of 0.05 and a power of 80\%.

Table~\ref{table:lab_demo} shows the characteristic information of the participants.
Forty participants (N=40) were enrolled (F=23, M=16, NB=1), aged between 18 to 65+ years, with 92.5\% falling in the 18-34 age range. 
Regarding online health information-seeking experience, 22.5\% frequently or always used online sources, 62.5\% occasionally searched online, and 15.0\% rarely used online resources. 
For the frequency of AI usage, 60.0\% frequently or always used AI tools, 22.5\% occasionally used AI, and 17.5\% rarely or never used AI.


\subsubsection{Study Procedure}
Each participant was invited to the institute for a single session at the lab. The researcher first provided an overview of the study and task details, after which participants gave informed consent before the lab session. During the pre-survey, participants provided their demographic information (age, gender, occupation) and their experiences with online health information search and interactions with AI.

Upon completing the pre-survey and successfully calibrating the sensors, participants began the formal reading task. During the reading task, each participant reviewed 12 sets of health information: six were labeled as from human professionals and six as from AI, regardless of the actual source. Sources and labels were counterbalanced to minimize order effects. The entire session lasted approximately 60 minutes, and participants were rewarded with \texteuro10 for participating. 
The study received approval from our institute's ethics and data protection committee.
The procedure of the lab study is detailed in Fig~\ref{fig:procedure}(b).


\subsubsection{Data Pre-processing}

\textit{Self-reported Trust Scores.}
To assess how factors such as information source, labeling, and information type affect trust in online health information, we first checked the suitability of the data for statistical analysis. A Shapiro-Wilk test~\cite{SHAPIRO1965} confirmed that the self-reported trust scores deviated from a normal distribution. Therefore, we applied generalized estimating equations (GEE)~\cite{gee} to analyze trust differences across information sources and labels, because of its robustness to violations of normality and flexibility to handle repeated ordinal measures.
Additionally, we conducted Spearman correlation analyses~\cite{spearman} \revise{with Bonferroni corrections to explore relationships among the variables.}
\revise{Consistent with Study 1, and given that the GEE results (Table~\ref{tab:lab_result_trust}) indicated no significant interaction effects between the independent variables of source and labeling, we averaged the repeated measures for each participant into a single observation across conditions. This simplification allowed us to focus on the key exploratory relationships while maintaining analytical clarity.}

\textit{Eye Tracking Data Processing.}
Raw eye-tracking data were extracted from the eye tracker (Tobii Pro Fusion) using Tobii Pro Lab software~\cite{TobiiProLab}), and time-synchronized with stimuli. 
As shown in Fig~\ref{fig:stimuli-lab-study} (Top), there are three Areas of Interest (AOIs): AOI-1 (disclosed label of source), AOI-2 (health information), and AOI-3 (rating scale). 
We chose fixation threshold of 30$^\circ$ for velocity and 60 ms for duration, as suggested by online information reading task~\cite{van2011defining}. Gaze features including gaze duration, fixation (count and duration), saccade (count and length), and pupil diameter were calculated to understand how users read the information. 
We excluded participants whose gaze accuracy fell below 90\%, resulting in 38 participants' eye-tracking data being retained for further analysis.
After transforming data through Aligned Ranked Transformation (ART)~\cite{art_transformation}, we confirmed the non-normality of eye tracking data with the Shapiro-Wilk test.

\textit{Physiological Signal Processing.}
Physiological signal was processed using Vsrrp98 software (v13.1.4)~\cite{physio_devices}, following the practise in prior research \cite{eda_practice, Ahmad_Muneeb}.
Key physiological features derived from the raw ECG data using interval-to-interval window size included BPM, BPI, and the main HRV metrics - the root mean square of successive differences (RMSSD). For EDA data, we used the continuous decomposition analysis method~\cite{benedek2010continuous} to separate it into the tonic SCL and phasic SCR components, then calculated the mean SCL and SCR values, SCR count. Skin temperature readings were screened for any abnormal responses.
We excluded SCL and SCR data when more than 4 out of 12 stimuli have values lower than .01$\mu$S or exceeded 50$\mu$S, as these readings likely resulted from loss recording or movement artifacts.
As a result, we retained data from 34 participants for SCL and SCR analysis, and 40 participants for ECG and skin temperature analysis.
Following preprocessing, we used the Shapiro-Wilk test to assess normality, revealing that all physiological features were not normally distributed. 

\revise{Given the exploratory nature of our investigation and the presence of multiple comparisons, we applied appropriate corrections based on the type of data. First, self-reported data were analyzed using a single GEE test, thus no multiple comparison correction was necessary. Second, for eye-tracking data, where multiple tests were conducted for different features, we applied False Discovery Rate (FDR) correction~\cite{fdr_bh} to control for potential inflation of Type I errors. Third, for physiological data, no multiple comparison correction was applied because most of the physiological features (e.g., RMSSD, ECG, EDA) were uncorrelated, as confirmed by correlation analysis, and each feature was analyzed independently. This approach reflects our goal of treating these features as distinct, non-overlapping measures, without assuming that they influence each other.}

\subsection{Findings}

\subsubsection{Descriptive Statistics}

As shown in Table~\ref{tab:lab_descriptive}, participants demonstrated a generally positive attitude toward technology, with an average trust in technology score of 3.36 (SD=.23). Their eHealth literacy averaged 3.69 (SD=.25), indicating proficiency in searching for digital health information. AI literacy was even higher, with an average score of 3.78 (SD=.20), suggesting a strong understanding of AI and its applications. 

\begin{table}[!htbp]
\centering
\renewcommand{\arraystretch}{1.20}
\scriptsize
\setlength{\tabcolsep}{3pt} 
\begin{tabular*}{\textwidth}{@{\extracolsep{\fill}}p{2cm}p{6cm}p{2cm}p{2cm}@{}}
\hline
\textbf{} & \textbf{Measures} & \textbf{Mean} & \textbf{SD} \\ \hline

\multirow{6}{*}{Pre-survey} 
 & Familiarity of AI & 3.58 & .96 \\ \cline{2-4} 
 & Perceived Reliability of AI & 3.13 & .61 \\ \cline{2-4} 
 & Perceived Reliability of Human Professionals & 3.78 & .53 \\ \cline{2-4}  
 & Propensity of Trust (PPT) & 3.54 & .33 \\ \cline{2-4}  
 & eHealth literacy & 3.69 & .25 \\ \cline{2-4}  
 & AI literacy & 3.78 & .20 \\ 

\hline

\textbf{} & \textbf{Conditions} & \textbf{Mean} & \textbf{SD} \\ \hline

\multirow{8}{*}{Trust score} 
 & Source (Human) \& Label (Human) & 3.67 & .63 \\ \cline{2-4} 
 & Source (Human) \& Label (AI) & 3.56 & .64 \\ \cline{2-4}
 & Source (LLM) \& Label (Human) & 3.92 & .56 \\ \cline{2-4} 
 & Source (LLM) \& Label (AI) & 3.78 & .63 \\ \cline{2-4} 
 & Source (Human), regardless of Label & 3.62 & .64 \\ \cline{2-4} 
 & Source (LLM), regardless of Label & 3.85 & .60 \\ \cline{2-4} 
 & Label (Human), regardless of Source & 3.80 & .61 \\ \cline{2-4} 
 & Label (AI), regardless of Source & 3.67 & .65 \\ 
\bottomrule
\end{tabular*}
\caption{Descriptive statistics of the lab study.}
\label{tab:lab_descriptive}
\vspace{-3mm}
\end{table}

Regarding the perceived trust, \major{the lab study results closely mirrored those of the online survey, despite being based on separate participant samples and independently collected data.}
The \revise{self-reported} trust scores from the lab study varied depending on both the source and the labeling of the health information. 
Information both sourced from and labeled as from human professionals had an average trust score of 3.67 (SD=.63). When human-sourced information was labeled as AI, the score slightly decreased to 3.56 (SD=.64). LLM-sourced information labeled as from human received the highest trust score of 3.92 (SD=.56), while information sourced from LLM and labeled as from AI had a trust score of 3.78 (SD=.63). 
Overall, participants reported higher trust in LLM-sourced information (M=3.85, SD=.60) than in human-sourced information (M=3.62, SD=.64), \major{echoing the trend observed in the online survey and} indicating a growing acceptance of AI (i.e., LLM) in health contexts. 
However, information labeled as coming from human professionals was trusted more (M=3.80, SD=.61) than that labeled as AI (M=3.67, SD=.65), suggesting that \major{labeling plays an influential role in trust formation, potentially even more than the actual source.}
\major{These findings reinforced the patterns found in the online survey while providing additional validity through the lab sessions.}


\subsubsection{Analysis of Self-reported Trust}

\begin{table}[!ht]
\centering
\renewcommand{\arraystretch}{1.20}
\scriptsize
\setlength{\tabcolsep}{2pt} 
\begin{tabular*}{\textwidth}{@{\extracolsep{\fill}}p{1.8cm}p{3.8cm}p{1.8cm}p{1.5cm}p{1.8cm}p{1.0cm}@{}}
\hline
\textbf{Outcomes} & \textbf{Conditions} & \textbf{Coefficient} & \textbf{P-value} & \textbf{Effect ($Std. \beta$)} & \textbf{Sig} \\ \hline
\multirow{3}{*}{Trust score} 
 & Source (Human vs. LLM) & .22 & .00 & .35 (medium) & ** \\ \cline{2-6} 
 & Label (Human vs. AI) & -.15 & .01 & .23 (medium) & ** \\ \cline{2-6} 
 & Source $*$ Label & .03 & .71 & .05 (small) &  \\ 

\bottomrule
\end{tabular*}
\caption{Results from the GEE analysis on the self-reported trust score. (**$p$<.01, *$p$<.05)}
\label{tab:lab_result_trust}
\vspace{-3mm}
\end{table}

\begin{figure}[htbp]
\centering
\includegraphics[width=0.99\textwidth]{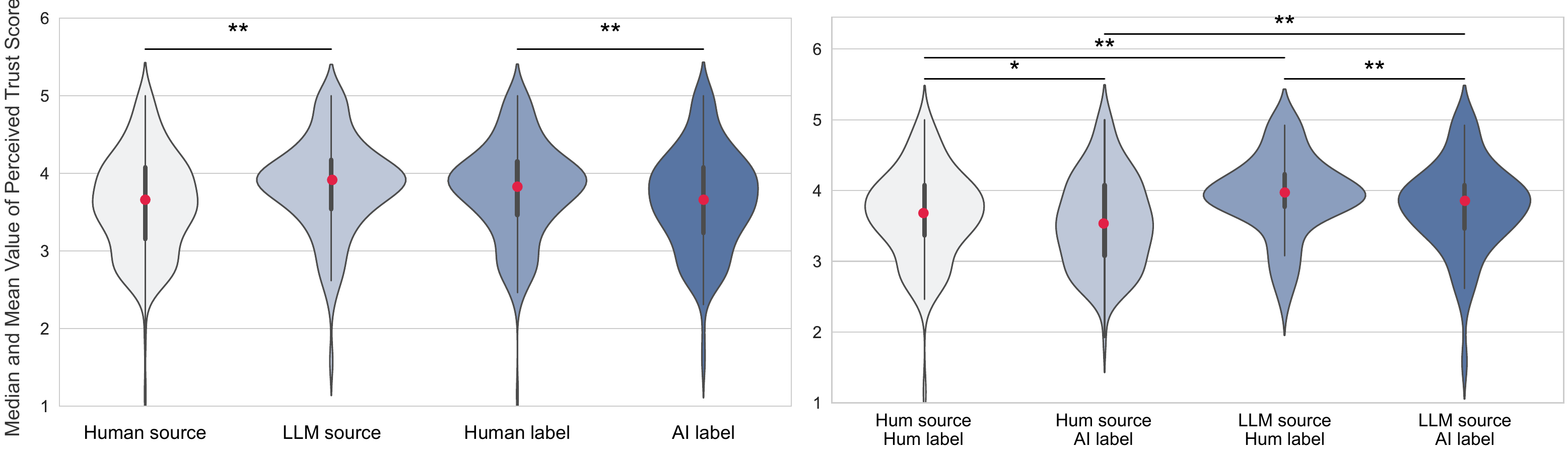}
\caption{\textbf{Left:} Perceived trust score in information by sources regardless of labels, and by 
labels regardless of sources.
\textbf{Right:} Perceived trust score based on different source and label conditions.
\major{
Each plot shows score density (width), with the red dot as the mean, the black line as the median, and thick bars denoting the interquartile range (IQR). Horizontal lines indicate significant pairwise differences (**$p$<.01, *$p$<.05).}
}
\vspace{-1.0mm}
\label{fig:lab_trust_analysis}
\end{figure}

Table~\ref{tab:lab_result_trust} presents the results from the GEE analysis on self-reported trust scores \major{from the lab study}. 
Consistent with the online survey, both the source and the label significantly impacted trust perceptions.
\major{Fig~\ref{fig:lab_trust_analysis} further illustrates the same pattern, echoing the online survey results. Trust was highest for LLM-sourced information labeled as human and lowest for human-sourced information labeled as AI.}

The analyses first revealed a significant effect of information source on trust, with a coefficient of 0.22 ($p<0.01$), indicating that LLM-sourced information was generally trusted more than human-sourced information, i.e., without knowledge of the actual source. This suggested that the source of information is crucial in shaping trust, as AI-generated content may be perceived as more structured and objective than human-authored content.
\revise{While the raw coefficient represented a modest change of 0.22 points on a 5-point Likert scale, the corresponding effect size ($Std. \beta$=0.35) was classified as medium. This reflected the bounded nature of the Likert scale, where even small raw differences can indicate meaningful relationships due to the relatively low variability in responses. Thus, the medium effect size underscored the practical relevance of the findings despite the small-scale differences.}

\begin{table}[!htbp]
\centering
\renewcommand{\arraystretch}{1.16}
\scriptsize
\setlength{\tabcolsep}{0pt} 
\begin{tabular*}{\textwidth}{@{\extracolsep{\fill}}p{1.2cm}p{0.6cm}p{1.4cm}p{1.0cm}p{1.2cm}p{1.2cm}p{1.9cm}p{0.6cm}@{}}
\hline

\textbf{Gaze} & \textbf{AOI} & \textbf{Condition} & \textbf{Coeff} & \textbf{\revise{P (Orig)}} & \textbf{\revise{P (Corrected)}} & \textbf{Effect ($Std. \beta$)} & \textbf{Sig} \\ 
\hline

\multirow{9}{*}{\parbox{1.2cm}{Fixation Count}} 
    & \multirow{3}{*}{AOI-1} 
    & Source & -4.02 & \textbf{.033} & \textbf{.079} & .22 (medium) & - \\ \cline{3-8} 
    & & Label & -4.54 & \textbf{.015} & \textbf{.055} & .25 (medium) & - \\ \cline{3-8} 
    & & Src $\times$ Lab & 6.01 & .082 & .150 & .33 (medium) &  \\ \cline{2-8}
    
    & \multirow{3}{*}{AOI-2} 
    & Source & 0.61 & .962 & .962 & .00 (small) &  \\ \cline{3-8} 
    & & Label & -3.23 & .827 & .910 & .02 (small) &   \\ \cline{3-8} 
    & & Src $\times$ Lab & 34.61 & \textbf{.036} & \textbf{.079} & .26 (medium) &  - \\ \cline{2-8}
    
    & \multirow{3}{*}{AOI-3} 
    & Source & -8.14 & .198 & .272 & .14 (small) &   \\ \cline{3-8} 
    & & Label & 9.90 & .151 & .237 & .17 (small) &   \\ \cline{3-8} 
    & & Src $\times$ Lab & -5.53 & .559 & .683 & .10 (small) &   \\ \hline

\multirow{9}{*}{\parbox{1.8cm}{Fixation \\Duration}} 
    & \multirow{3}{*}{AOI-1} 
    & Source & -51.39 & \textbf{.000} & \textbf{.000} & .43 (large) &  ** \\ \cline{3-8} 
    & & Label & -37.64 & \textbf{.006} & \textbf{.017} & .31 (medium) & * \\ \cline{3-8} 
    & & Src $\times$ Lab & 71.27 & \textbf{.000} & \textbf{.000} & .59 (large) & ** \\ \cline{2-8}
    
    & \multirow{3}{*}{AOI-2} 
    & Source & 4.60 & \textbf{.038} & \textbf{.069} & .14 (medium) & - \\ \cline{3-8} 
    & & Label & -2.80 & .218 & .343 & .09 (small) &   \\ \cline{3-8} 
    & & Src $\times$ Lab & 1.75 & .584 & .642 & .05 (small) &   \\ \cline{2-8}
    
    & \multirow{3}{*}{AOI-3} 
    & Source & -0.46 & .879 & .879 & .01 (small) &   \\ \cline{3-8} 
    & & Label & -1.61 & .553 & .642 & .05 (small) &   \\ \cline{3-8} 
    & & Src $\times$ Lab & 2.85 & .519 & .642 & .09 (small) &   \\ \hline

\multirow{9}{*}{\parbox{1.8cm}{Saccade Count}} 
    & \multirow{3}{*}{AOI-1} 
    & Source & -5.13 & \textbf{.044} & \textbf{.086} & .21 (medium) & - \\ \cline{3-8} 
    & & Label & -7.38 & \textbf{.013} & \textbf{.047} & .26 (medium) & * \\ \cline{3-8} 
    & & Src $\times$ Lab & 8.89 & \textbf{.047} & \textbf{.086} & .36 (medium) & - \\ \cline{2-8}
    
    & \multirow{3}{*}{AOI-2} 
    & Source & -4.35 & .787 & .787 & .03 (small) &   \\ \cline{3-8} 
    & & Label & -7.82 & .651 & .716 & .05 (small) &   \\ \cline{3-8} 
    & & Src $\times$ Lab & 43.77 & \textbf{.026} & \textbf{.071} & .28 (medium) & - \\ \cline{2-8}
    
    & \multirow{3}{*}{AOI-3} 
    & Source & -9.40 & .223 & .307 & .12 (medium) &   \\ \cline{3-8} 
    & & Label & 10.71 & .160 & .251 & .14 (medium) &  \\ \cline{3-8} 
    & & Src $\times$ Lab & -7.96 & .487 & .595 & .10 (medium) &   \\ \hline

\multirow{9}{*}{\parbox{1.8cm}{Saccade Length}} 
    & \multirow{3}{*}{AOI-1} 
    & Source & 0.03 & .124 & .341 & .17 (medium) &   \\ \cline{3-8} 
    & & Label & 0.01 &.531 & .649 & .08 (small) &   \\ \cline{3-8} 
    & & Src $\times$ Lab & -0.07 & \textbf{.020} & \textbf{.073} & .38 (medium) & - \\ \cline{2-8}
    
    & \multirow{3}{*}{AOI-2} 
    & Source & 0.00 & .355 & .558 & .08 (small) &   \\ \cline{3-8} 
    & & Label & 0.00 & .181 & .398 & .12 (medium) &   \\ \cline{3-8} 
    & & Src $\times$ Lab & 0.00 & .829 & .829 & .02 (small) &   \\ \cline{2-8}
    
    & \multirow{3}{*}{AOI-3} 
    & Source & 0.00 & .276 & .506 & .08 (small) &   \\ \cline{3-8} 
    & & Label & 0.00 & .456 & .627 & .06 (small) &   \\ \cline{3-8} 
    & & Src $\times$ Lab & 0.00 & .685 & .754 & .05 (small) &   \\ \hline

\multirow{9}{*}{\parbox{1.8cm}{Pupil\\Diameter\\Fixation}} 
    & \multirow{3}{*}{AOI-1} 
    & Source & -0.41 & \textbf{.002} & \textbf{.011} & .35 (medium) & * \\ \cline{3-8} 
    & & Label & -0.33 & \textbf{.018} & \textbf{.040} & .29 (medium) & * \\ \cline{3-8} 
    & & Src $\times$ Lab & 0.50 & \textbf{.003} & \textbf{.011} & .43 (large) &  * \\ \cline{2-8}
    
    & \multirow{3}{*}{AOI-2} 
    & Source & 0.00 & .673 & .823 & .01 (small) &   \\ \cline{3-8} 
    & & Label & 0.01 & .445 & .699 & .02 (small) &   \\ \cline{3-8} 
    & & Src $\times$ Lab & 0.00 & .898 & .932 & .00 (small) &   \\ \cline{2-8}
    
    & \multirow{3}{*}{AOI-3} 
    & Source & 0.00 & .227 & .416 & .03 (small) &   \\ \cline{3-8} 
    & & Label & 0.01 & .531 & .730 & .01 (small) &   \\ \cline{3-8} 
    & & Src $\times$ Lab & 0.00 & .932 & .932 & .00 (small) &   \\ 
\bottomrule    
\end{tabular*}
\caption{Results from the GEE analysis \revise{with False Discovery Rate (FDR) correction} on the eye tracking data. (**$p$<.01, *$p$<.05, -$p$<.10)}
\label{tab:gaze-results}
\vspace{-3.2mm}
\end{table}

Labeling also significantly impacted trust, with a coefficient of -0.15 ($p=0.01$), meaning information labeled as AI was trusted less than when labeled as human professionals. The negative coefficient suggested a preference for human-labeled information, as participants may associate human endorsement with greater credibility.
\revise{Similarly, while the raw change (-0.15) was modest, the standardized effect size ($Std. \beta$=0.23) reflected a medium effect, emphasizing that the impact of labeling, though subtle on the scale, has measurable and meaningful implications for trust perception.}

\major{Notably,} the interaction between source and label was not significant ($coefficient = 0.03, p = 0.71$), indicating that the combined influence of source and label does not affect trust beyond their individual effects. 
The small standardized effect size ($Std. \beta$=0.05) confirmed that this interaction effect is negligible.


\subsubsection{Analysis of Eye Movement Data}

The results of GEE analysis~\cite{gee} on eye tracking data are detailed in Table~\ref{tab:gaze-results}, showing varied eye movement patterns. 
\major{In AOI-1 (label area), fixation duration and pupil diameter of fixation showed significant differences by information sources and labels, while saccade count showed significant differences by information labels only.}

\begin{figure}[!ht]
\centering
\includegraphics[width=0.96\textwidth]{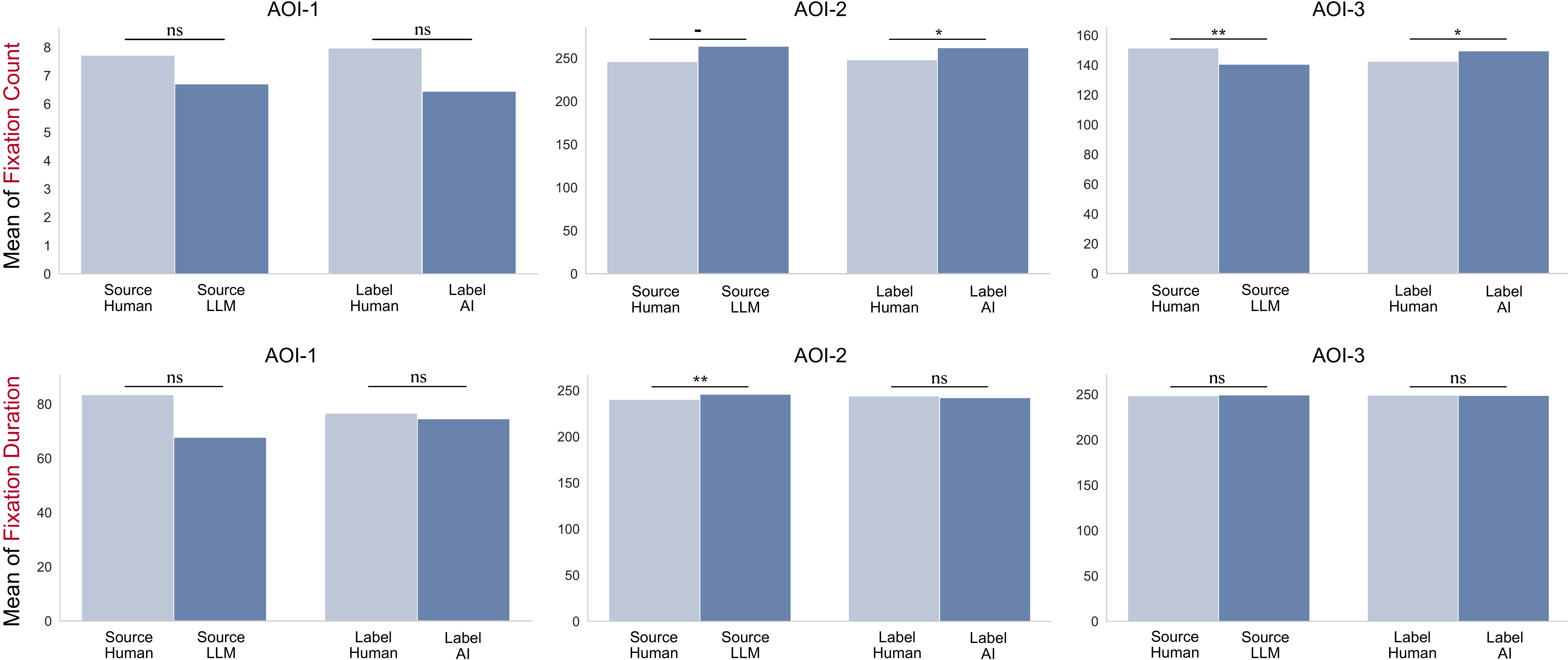}
\caption{Posthoc pairwise comparison by Wilcoxon signed-rank test with False Discovery Rate (FDR) correction of fixation features (count and duration) in three AOIs. (**$p$<.01, *$p$<.05, -$p$<.10, \major{``ns'' is not significant}).
}
\vspace{-1mm}
\label{fig:fixation}
\end{figure}

\begin{figure}[!ht]
\centering
\includegraphics[width=0.96\textwidth]{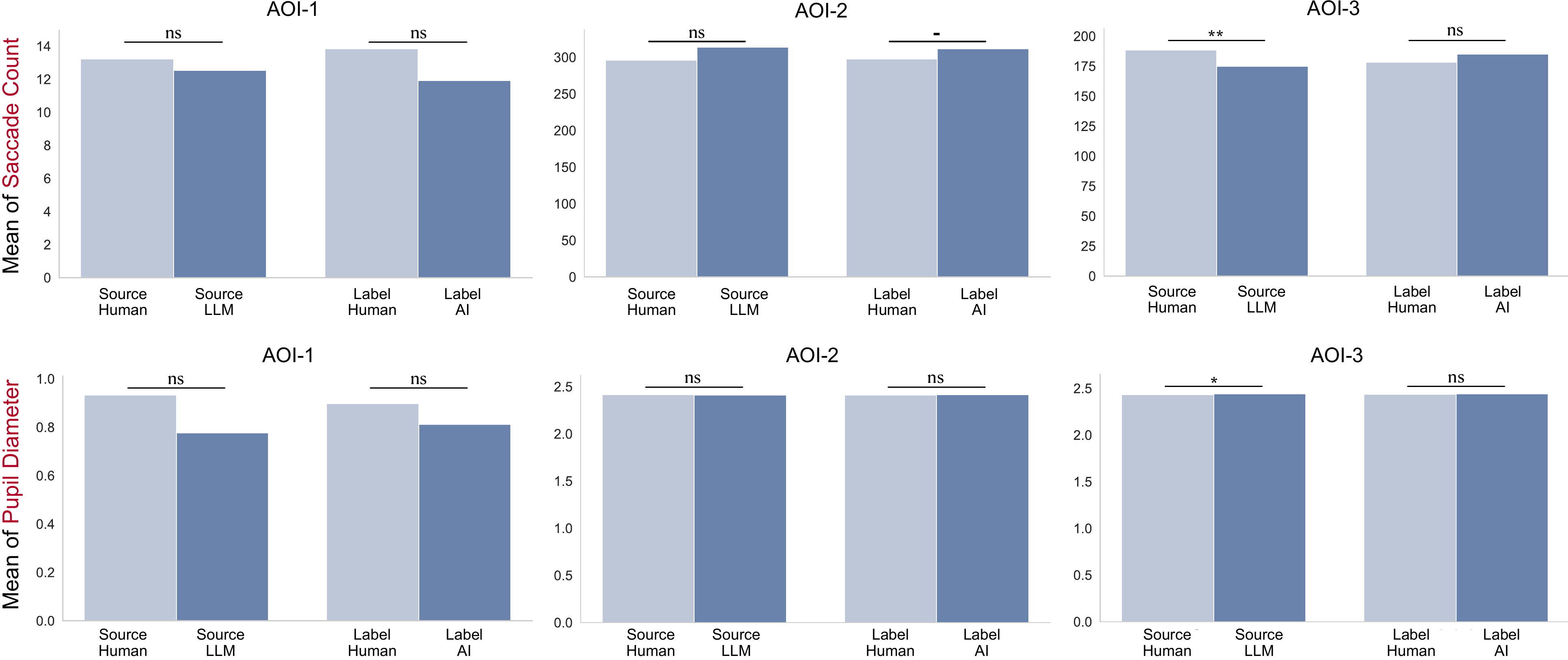}
\caption{Posthoc pairwise comparison by Wilcoxon signed-rank test \major{with False Discovery Rate (FDR) correction} of \textcolor{black}{saccade count} and \textcolor{black}{pupil diameter of fixation} in three AOIs. (**$p$<.01, *$p$<.05, -$p$<.10, \major{``ns'' is not significant}).}
\vspace{-2mm}
\label{fig:saccade_pupil}
\end{figure}

The post hoc comparisons shown in Fig~\ref{fig:fixation} and Fig~\ref{fig:saccade_pupil}, 
participants demonstrated higher \major{fixation counts ($p<.05$) and saccade counts ($p<.1$)} in AOI-2 (main health information area) under the AI label condition, indicating that participants assessed the information focusing more on the content itself rather than the label when they were informed that the information is from AI. 
This implies that \revise{trust-related} judgments in AI-labeled information were driven more by the actual content than the labeling of the source.
Participants also showed significantly fewer fixation counts ($p<.05$) in AOI-3 (rating area) under the human label condition compared to the AI label condition. This suggests that human labels might inspire greater confidence, potentially influencing how users \revise{rate the trust score} on the information.
When information was actually sourced from LLM, participants showed higher fixation duration ($p<.01$) and counts ($p<.1$) in AOI-2, suggesting a more careful reading of AI-generated content. 
Conversely, human-sourced information led to higher fixation and saccade counts in AOI-3 ($p<.01$), indicating that LLM-sourced information might inspire greater confidence, potentially influencing how users \revise{rate the trust score}, which aligns with the self-reported trust perceptions.


\subsubsection{Analysis of Physiological Signals}

Table~\ref{tab:physio-results} presents the results from GEE analysis on physiological data, shedding light on how physiological responses vary with different information sources and labeling.

\begin{table}[!htbp]
\centering
\renewcommand{\arraystretch}{1.02}
\scriptsize
\setlength{\tabcolsep}{1pt} 
\begin{tabular*}{\textwidth}{@{\extracolsep{\fill}}p{1.8cm}p{1.5cm}p{3.7cm}p{1.3cm}p{1.3cm}p{2.4cm}p{0.6cm}@{}}
\hline

\textbf{Outcomes} & \textbf{Features} & \textbf{Conditions} & \textbf{Coeff} & \textbf{P-value} & \textbf{Effect ($Std. \beta$)} & \textbf{Sig} \\ 
\hline

\multirow{9}{*}{\parbox{1.8cm}{ECG}} 
    & \multirow{3}{*}{BPM} 
    & Source (Human vs. LLM) & -0.58 & .571 & .07 (small) &  \\ \cline{3-7} 
    & & Label (Human vs. AI) & -1.10 & .288 & .13 (medium) &   \\ \cline{3-7} 
    & & Source $\times$ Label & 1.38 & .341 & .17 (medium) &  \\ \cline{2-7}
    
    & \multirow{3}{*}{RMSSD} 
    & Source (Human vs. LLM) & 2.11 & .435 & .12 (medium) &  \\ \cline{3-7} 
    & & Label (Human vs. AI) & 5.21 & .025 & .29 (medium) & * \\ \cline{3-7} 
    & & Source $\times$ Label & -4.45 & .179 & .25 (medium) &  \\ \cline{2-7}
    
    & \multirow{3}{*}{BPI} 
    & Source (Human vs. LLM) & 8.88 & .242 & .12 (medium) &  \\ \cline{3-7} 
    & & Label (Human vs. AI) & 10.43 & .225 & .14 (medium) &  \\ \cline{3-7} 
    & & Source $\times$ Label & -17.10 & .153 & .24 (medium) &  \\ \hline

\multirow{6}{*}{\parbox{1.8cm}{EDA}} 
    & \multirow{3}{*}{SCL} 
    & Source (Human vs. LLM) & 0.03 & .949 & .04 (small) &  \\ \cline{3-7} 
    & & Label (Human vs. AI) & -0.77 & .061 & .12 (medium) & - \\ \cline{3-7} 
    & & Source $\times$ Label & 0.38 & .414 & .06 (small) &  \\ \cline{2-7}
    
    & \multirow{3}{*}{SCR} 
    & Source (Human vs. LLM) & -0.56 & .399 & .05 (small) &  \\ \cline{3-7} 
    & & Label (Human vs. AI) & -0.92 & .082 & .08 (small) & - \\ \cline{3-7} 
    & & Source $\times$ Label & -0.98 & .576 & .08 (small) &  \\ \hline

\multirow{3}{*}{\parbox{1.8cm}{Temperature}} 
    & \multirow{3}{*}{---} 
    & Source (Human vs. LLM) & 0.46 & .022 & .31 (medium) & * \\ \cline{3-7} 
    & & Label (Human vs. AI) & 0.42 & .029 & .28 (medium) &* \\ \cline{3-7} 
    & & Source $\times$ Label & -0.57 & .058 & .39 (medium) & - \\  

\bottomrule    
\end{tabular*}
\vspace{-2.0mm}
\caption{Results from GEE analysis on physiological signals. (**$p$<.01, *$p$<.05, -$p$<.10)}
\label{tab:physio-results}
\vspace{-3.6mm}
\end{table}

\begin{figure}[!ht]
\centering
\includegraphics[width=0.999\textwidth]{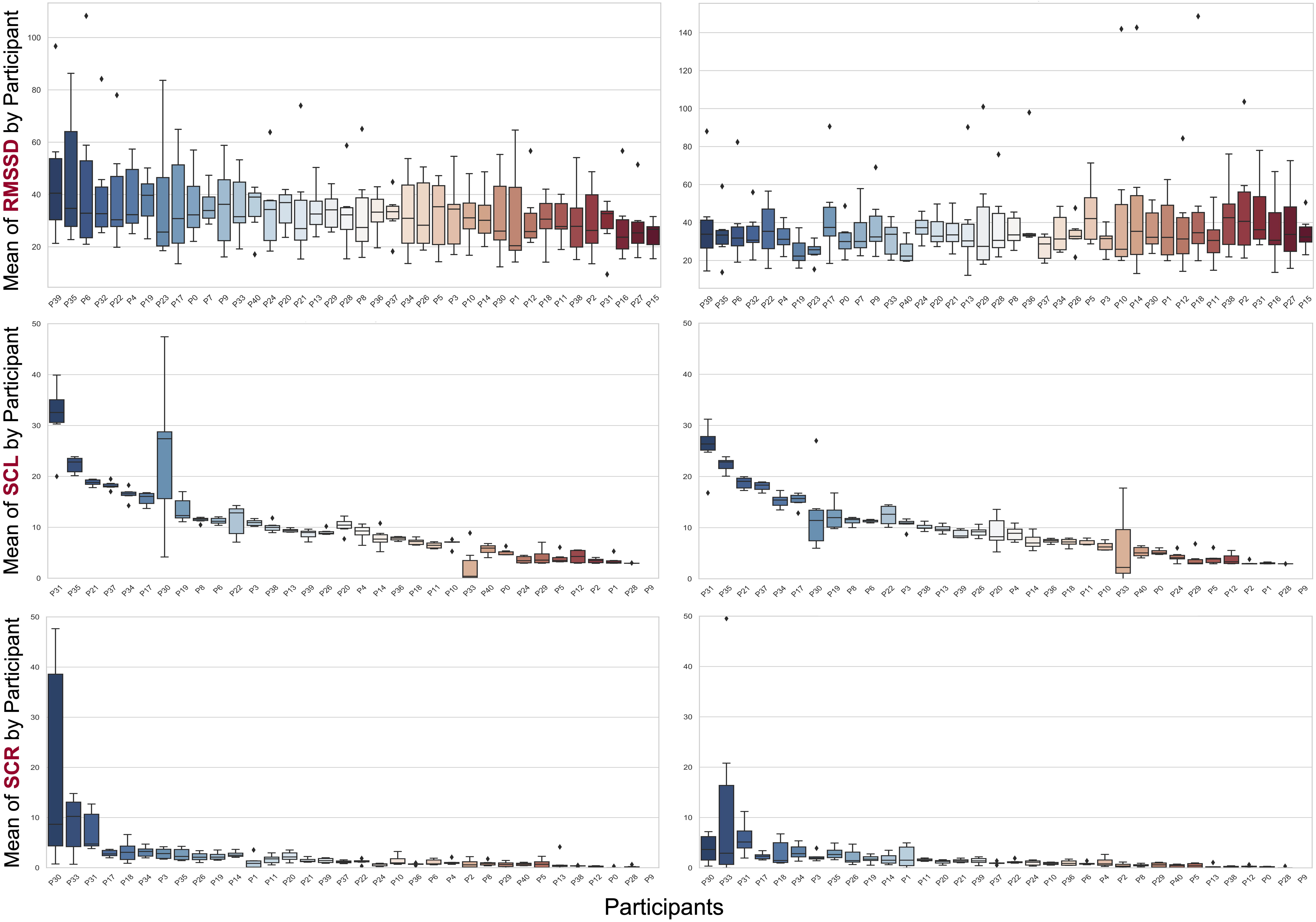}
\vspace{-3.0mm}
\vspace{-3.0mm}
\caption{
Pairwise comparison \revise{without correction} on features of RMSSD and SCR per participant. 
Each boxplot shows the distribution (median, IQR, outliers) for each participant under two labeling conditions:
participants read the information labeled as from ``Human Professionals'' \textbf{(Left)} and from ``AI'' \textbf{(Right)}, regardless of the source. 
\major{Color gradient reflects participant-wise ordering based on decreasing RMSSD or SCR values to facilitate visual comparison; the color itself carries no semantic meaning.}
}
\label{fig:physio}
\vspace{-4.6mm}
\end{figure}

\revise{RMSSD, a feature derived from ECG data, was significantly higher for AI-labeled information compared to human-labeled information ($p=.025$). Higher RMSSD indicates greater heart rate variability (HRV), which is often associated with lower physiological arousal.}
This aligns with the \major{gaze patterns that participants paid less attention to labeling area (AOI-1) under ``AI'' labels than ``Human'' labels, as indicated by significantly reduced fixation duration, saccade count, and pupil diameter (see Table~\ref{tab:gaze-results}).}

Skin temperature responses also varied significantly between human and AI labels ($p=.029$), as well as between human and LLM sources ($p=.022$). Higher skin temperature in response to AI labels and sources suggests participants may have experienced increased emotional arousal or stress when interacting with AI-associated content.

SCL ($p=.061$) and SCR ($p=.082$) average values did not exhibit statistically significant differences, as shown in Fig~\ref{fig:physio}. 
This suggests that EDA components, at least within our study, were not discriminative of physiological arousal when users encountered human versus AI-generated information.



\subsubsection{Correlation Analysis}
\begin{figure}[!ht]
\centering
\includegraphics[width=0.99\textwidth]{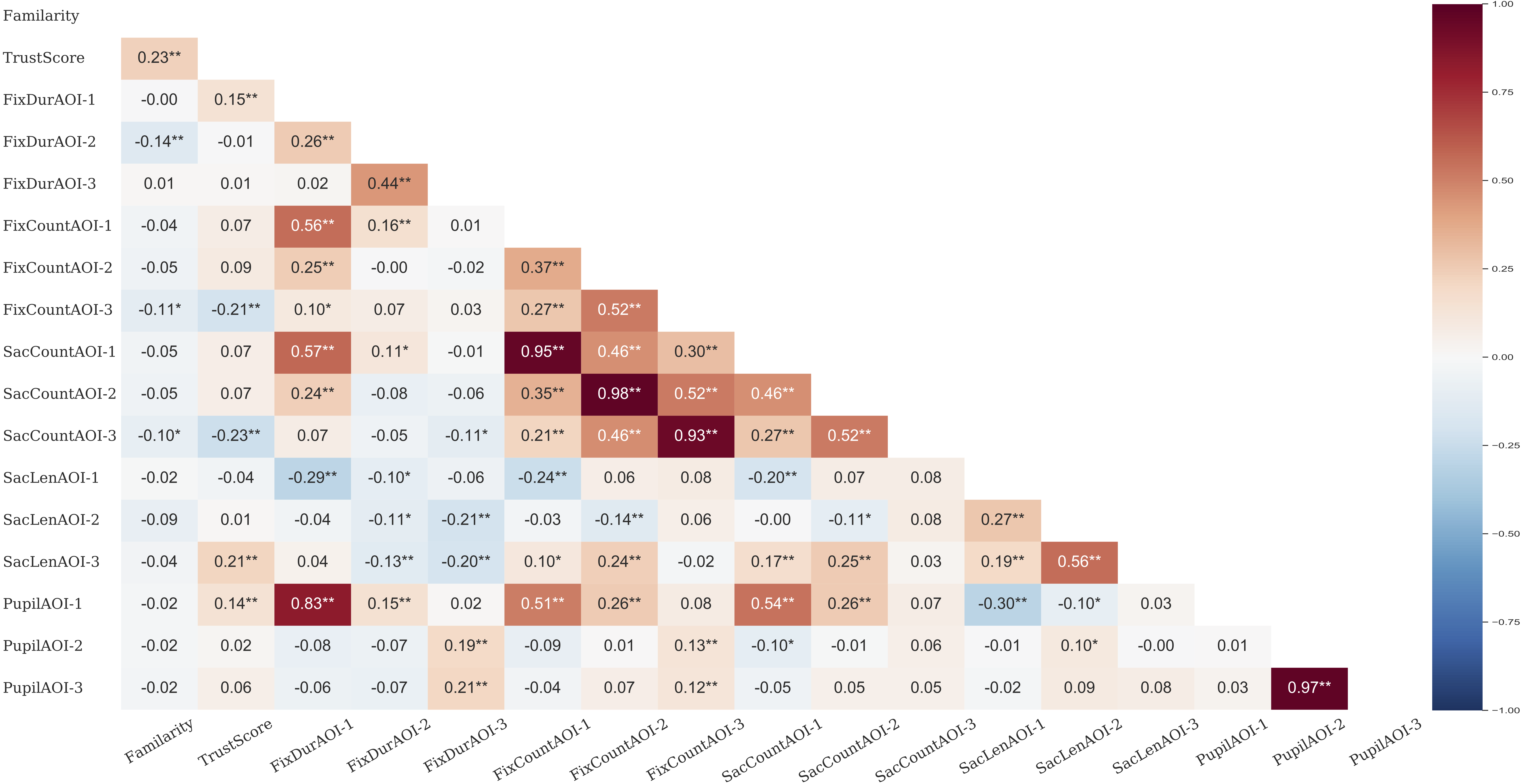}
\caption{Spearman correlation \revise{with Bonferroni corrections} between trust perceptions and the gaze features. (**$p$<.01, *$p$<.05).
Note: 
``FixDurAOI-'': fixation duration in AOI-; 
``FixCountAOI-'': fixation count in AOI-;
``SacCountAOI-'': saccade count in AOI-;
``SacLenAOI-'': saccade lenth in AOI-;
``PupilAOI-'': pupil diameter of fixation in AOI-.}
\label{fig:correlation_lab_gaze}
\vspace{-1mm}
\end{figure}

\begin{figure}[!ht]
\centering
\includegraphics[width=0.99\textwidth]{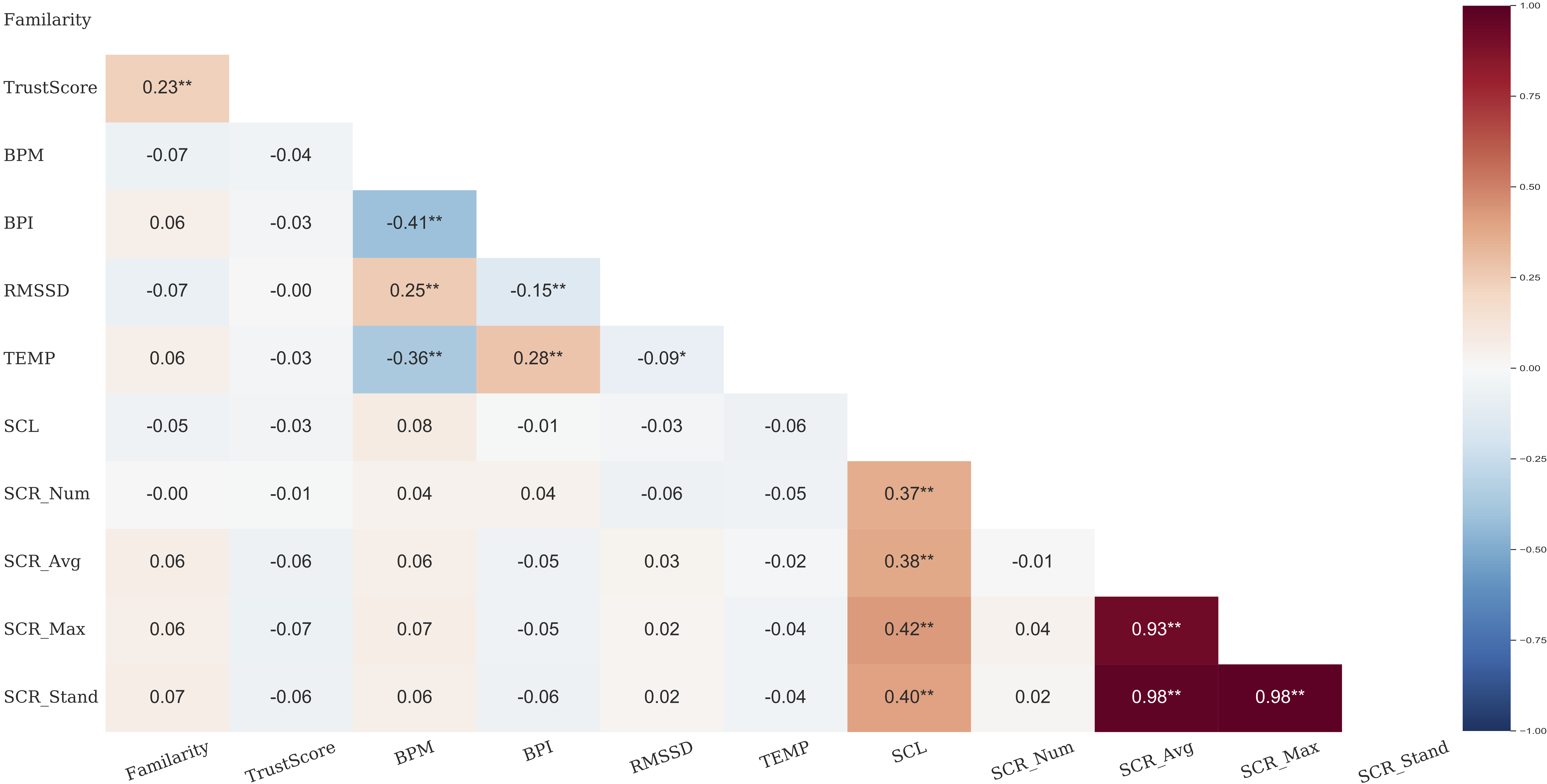}
\caption{Correlation on variables. (**$p$<.01, *$p$<.05).
Note: ``SCR\_Num'': number of SCR; ``SCR\_Avg'': average value of SCR; ``SCR\_Max'': maximum value of SCR; ``SCR\_Stand'': standard value of SCR.}
\label{fig:correlation_lab_physio}
\vspace{-2mm}
\end{figure}

The Spearman correlation analysis~\cite{spearman} in Fig~\ref{fig:correlation_lab_gaze} and ~\ref{fig:correlation_lab_physio} revealed significant relationships between \revise{the self-reported trust score} and various gaze and physiological features, indicating how participants' perceived trust in health information is linked to their behavioral and physiological responses.

Familiarity with the health question showed a strong positive correlation with trust in the information ($p<.01$). Among gaze features, fixation duration in AOI-1 (label area) positively correlated with the perceived \revise{trust score} ($p<.01$), indicating that higher trust levels are associated with a longer focus on the labeling of information sources. Additionally, pupil diameter of fixation in AOI-1 ($p<.01$) also correlated positively with trust score.
Fixation and saccade count in AOI-3 (rating area) were negatively correlated with trust, implying that participants who gave lower trust in the information exhibited more frequent saccadic movements in the rating area, likely reflecting efforts to evaluate or verify the information further.

No significant correlations were found between physiological features and trust levels. However, there were correlations observed among the physiological features themselves, such as BPM (Heartbeats), SCL, SCR, and skin temperature, though these did not directly link to trust.


\subsubsection{Predictions using Behavioral and Physiological Sensing}

To explore trust perception \revise{(i.e., self-reported trust scores)} through behavioral and physiological responses, we defined two tasks: 1) predicting participants' perceived trust score in health information and 2) classifying the source of the health information.

\begin{table}[!h]
\centering
\scriptsize
\renewcommand{\arraystretch}{1.1}
\setlength{\tabcolsep}{2pt} 
\begin{tabular*}{\textwidth}{@{\extracolsep{\fill}}p{3.5cm}p{1.2cm}p{1.2cm}p{1.2cm}p{1.2cm}p{1.2cm}p{1.2cm}@{}}
\hline

\multirow{2}{*}{\textbf{Models}} & \multicolumn{2}{l}{\textbf{Gaze Only}} & \multicolumn{2}{l}{\textbf{Physio Only}} & \multicolumn{2}{l}{\textbf{Gaze + Physio}} \\ \cline{2-7}

& MAE &  R$^2$ & MAE & R$^2$ & MAE &  R$^2$ \\ \hline


SVR & .29 & .06 &   .33 & .06 &   .28 & .10 \\ \cline{1-7} 
Linear Regression & .25 & .20 &   .31 & .01 &   .24 & .20 \\ \cline{1-7} 
Ridge Regression & .24 & .21 &   .31 & .01 &   .24 & .23 \\ \cline{1-7} 
Random-Forest Regression & .23 & .25 &   .25 & .19 &   .20 & .35 \\ \cline{1-7} 
XGBoost & .24 & .22 &   .28 & .08 &   .23 & .23 \\ \cline{1-7} 

\end{tabular*}%
\caption{Prediction on perceived trust score through regression using gaze and physiological features.}
\label{tab:lab_regression}
\vspace{-3mm}
\end{table}

\begin{figure}[!ht]
\centering
\includegraphics[width=0.9999\textwidth]{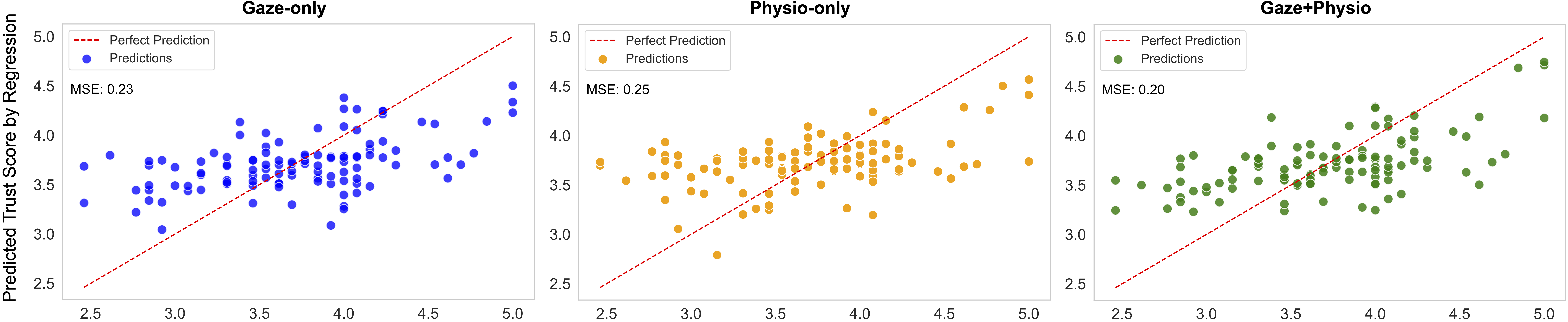}
\caption{Prediction of perceived trust score using the Random-Forest Regression model on three different features set: Gaze-only, Physiology-only, Gaze+Physiology.
\major{
Each dot represents one participant’s predicted vs. actual self-reported trust score, with the red dashed line indicating perfect prediction.}
}
\label{fig:lab_regression}
\end{figure}

For trust prediction, we first explored how regression models approximate perceived trust scores using regression models: linear regression (LR), ridge regression, SVM and random forest-based regressions, and XGBoost. 
As shown in Table~\ref{tab:lab_regression}, the random forest regressor on the combined Gaze+Physio feature set achieved the lowest MSE of .20 and highest $R^2=.35$ among the three feature sets, indicating the best performance. This highlights the value of combining gaze and physiological features for trust assessment. 
\major{Fig~\ref{fig:lab_regression} illustrates the regression results on three different feature sets.}

\begin{table}[!ht]
\centering
\scriptsize
\renewcommand{\arraystretch}{1.20}
\setlength{\tabcolsep}{2pt} 
\begin{tabular*}{\textwidth}{@{\extracolsep{\fill}}p{1.8cm}p{2.2cm}p{2.0cm}p{2.0cm}p{2.0cm}@{}}
\hline

\multirow{3}{*}{\textbf{Features}} & \multirow{3}{*}{\textbf{Models}} & \multicolumn{2}{c}{\textbf{Trust Level}} & \textbf{Source} \\ \cline{3-5}

 & & \textbf{2-class\newline(Acc / F1)} & \textbf{\revise{3-class\newline(Acc / F1)}} & \textbf{2-class\newline(Acc / F1)} \\ \hline


\multirow{10}{*}{\textbf{Gaze Only}} &
 LR & .65 / .62 & .57 / .57 & .62 / .55  \\ \cline{2-5} 
 & RF & .69 / .65 & .57 / .57 & .57 / .52  \\ \cline{2-5} 
 & SVM & .51 / .53 & .43 / .42 & .60 / .48  \\ \cline{2-5}
 & MLP & .57 / .58 & .32 / .32 & .44 / .53   \\ \cline{2-5}
 & 
 
 GradientBoost & .72 / .66 & .54 / .54 & .52 / .52 \\ \cline{2-5} 
 & AdaBoost & .67 / .64 & .58 / .58 & \underline{\textit{\textbf{.65}}} / .52 \\ \cline{2-5} 
 & XGBoost & .70 / .66 & .54 / .54 & .43 / .52 \\ \cline{2-5}
 & 
 
 Voting & \underline{\textit{\textbf{.73}}} / .67 & .54 / .54 & .60 / .49 \\ \cline{2-5}
 & 
 
 Stacking & .70 / .66 & \underline{\textit{\textbf{.59}}} / .58 & .49 / .55 \\ \cline{2-5}
 & 
 
 Bagging & .70 / .66 & .57 / .57 & .57 / .47 \\ \hline


\multirow{10}{*}{\textbf{Gaze + Physio}} &
 LR & .65 / .62 & .58 / .56 & .58 / .54 \\ \cline{2-5} 
 & RF & .69 / .65 & \underline{\textit{\textbf{.63}}} / .63 & .60 / .52  \\ \cline{2-5} 
 & SVM & .51 / .53 & .43 / .43  & .60 / .49 \\ \cline{2-5}
 & MLP & .53 / .60 & .48 / .47 & .59 / .50  \\ \cline{2-5}
 & 
 
 GradientBoost & \underline{\textit{\textbf{.72}}} / .68 & .59 / .56 & .53 / .53 \\ \cline{2-5} 
 & AdaBoost & .66 / .64 & .54 / .54 & \underline{\textit{\textbf{.65}}} / .64 \\ \cline{2-5} 
 & XGBoost & .65 / .67 & .57 / .57 & .57 / .52 \\ \cline{2-5}
 & 
 
 Voting & .67 / .67 & .59 / .58 & .60 / .53 \\ \cline{2-5}
 & 
 
 Stacking & .69 / .66 & .60 / .60 & .48 / .52 \\ \cline{2-5}
 & 
 
 Bagging & .70 / .66 & .61 / .61 & .54 / .53 \\ \hline

\end{tabular*}%
\caption{Classification on trust level (high, medium, low) and the source of information using gaze and physiological features.}
\label{tab:lab_classification}
\end{table}

Next, we performed \revise{both binary (i.e., high vs. low) and three-class (i.e., high vs. medium vs. low) classification of trust levels} based on participants' self-reported \revise{trust scores}. 
As shown in Table~\ref{tab:lab_classification}, the ensemble method (voting model) achieved the highest accuracy (0.73) for binary classification using gaze-only features, while random forest achieved the highest accuracy (0.63) for three-class classification using combined gaze-physiological features. Interestingly, combining gaze and physiological features did not improve performance across all models, for instance, the gradient boosting model achieved slightly lower accuracy (0.72) of binary classification when incorporating both feature sets compared to using gaze features alone.
\revise{These results indicate that gaze features alone achieved higher classification accuracy for binary trust levels compared to combined gaze and physiological features. It suggests that gaze features may play a more prominent role in predicting trust levels than physiological responses, at least in the context of this study.}

For the second task to classify the information source, combining gaze and physiological features yielded the best results. The AdaBoost model achieved the highest accuracy of 0.65 and F1 score of 0.64, indicating that physiological responses complement gaze features in distinguishing between human- and LLM-generated health information.

\begin{figure}[!htbp]
\centering
\includegraphics[width=0.9998\textwidth]{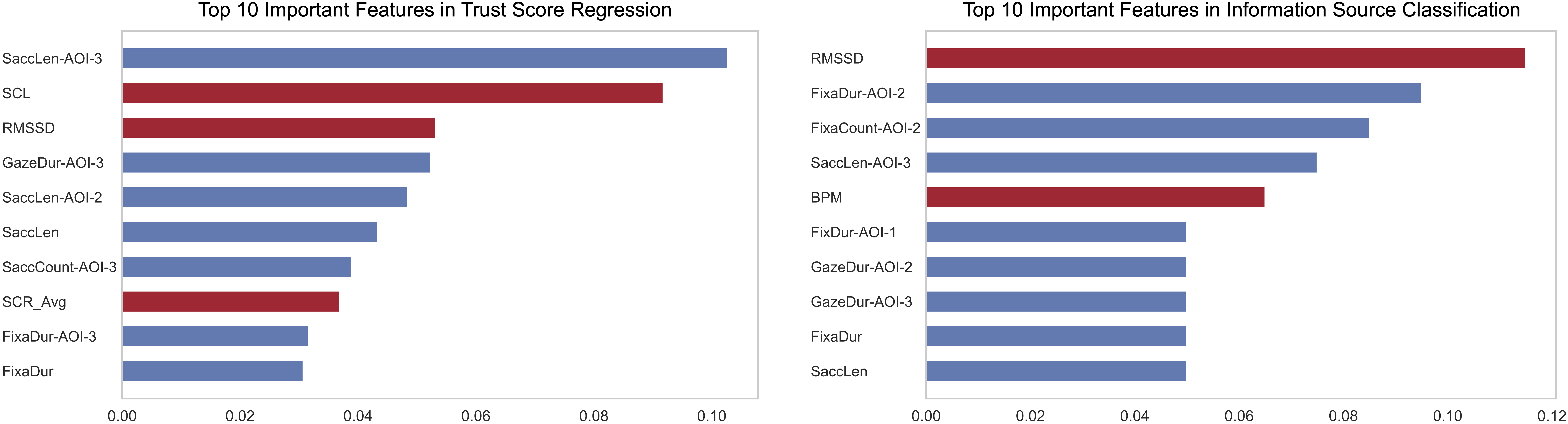}
\caption{
Top 10 important gaze and physiological features in Random Forest regressor for predicting perceived trust scores (\textbf{Left}) and in AdaBoost classifier for classifying the source of health information (\textbf{Right}), \major{based on SHAP values computed on the test set. \major{Blue bars represent gaze features; red bars represent physiological features.}}
}
\label{fig:lab_feature_importance}
\end{figure}

Fig~\ref{fig:lab_feature_importance} presents feature importance for the prediction tasks following \major{SHAP framework} proposed \major{by Lundberg and Lee~\cite{feature_importance}} for better interpreting the model prediction.
In summary, gaze features are effective for predicting trust perceptions, while combining gaze and physiological features could improve the classification of information sources. 
The robust performance of ensemble methods across both tasks highlights their potential in developing tools to assess trust-related responses in health communication by leveraging gaze and physiological signals.

\section{Discussion}



\major{We conducted an online survey and a lab study in this work to investigate how users' trust responds to human versus AI-generated content, and in what ways trust in online health information may be influenced by including transparency labels as simple as ``Human'' versus ``AI'' labels on personal health information.}
Our findings showed that self-reported trust in digital health information is influenced by its actual source and disclosed labeling of the source. Further, the impacts of these conditions were also evident at a behavioral and physiological level. 
Below, we discuss these aspects in detail.

\subsection{Users may Prefer \revise{LLM-Sourced} Health Information, but An AI Label Lowers Their Trust}

Both studies tested \textbf{(RQ1)} if the actual source, disclosed label, and type of information influence perceived trust in online personal health information. Our findings revealed that \revise{LLM-sourced} content is trusted more than human-sourced content, regardless of labeling, whereas human professional labels are trusted more than AI labels. 
Trust however remained consistent across different information types (general, symptom, or treatment-related), suggesting that the source and labeling, rather than the type of information, are the primary determinants of perceived trust.

The observed difference in trust perception was evident in both self-reported trust scores (i.e., higher trust scores of LLM-generated information) and qualitative data, which suggests that participants have perceived subtle distinctions of information presentation styles in the LLM- versus human professionals-sourced information that provided cues for trust. 
The stronger effect observed in Study 2 (lab study with the within-subjects design) compared to Study 1 (survey study with the between-subjects design) further supports this, as the within-subjects design allowed participants to compare responses from both sources side by side.
While we cannot conclusively determine the specific factors in information quality driving higher trust, our findings imply that LLM-generated content may convey an impression of clarity or objectivity that resonates more strongly with participants.
Our observation that \revise{LLM-sourced} information was trusted more than that from human professionals may reflect advancements in LLMs like ChatGPT, which can produce structured and high-quality responses~\cite{t10,t11}. 
\major{Notably, GPT-4 generated responses have been found to be perceived as more human-like than actual human-authored content and other studies find that LLM-generated content is often indistinguishable from human-generated text~\cite{llm_text_quality_1}.}
\major{This explanation (i.e., generally higher language quality of LLM-generated responses as a basis of trust) aligns with Dalton et al.'s~\cite{Dalton} proposal of emergent conversational information-seeking powered by LLMs,} and is evident when assessing how LLMs are being used in the context of healthcare~\cite{t6,t7,t10,t11,gg_gpt_2}. 

Furthermore, researchers suggest that people prefer algorithms to humans in certain tasks and it could relate to individuals' machine heuristic (rule of thumb that machines are more secure and trustworthy than humans~\cite{Logg, Sundar2019}\major{)}. 
In our studies, qualitative analyses (Sec \ref{sec:qual}) further confirmed that participants attributed the higher trust in LLM-generated content to its efficiency, capacity to process extensive health data~\cite{Singhal2023}, and objective language style~\cite{gg_gpt_1, gg_gpt_2}. 
This suggests that LLMs' (e.g., GPT-4~\cite{gpt4}) ability to deliver comprehensive and objective health information resonates with users, positioning them as reliable sources of health information.


\major{Paradoxically, when health information was labeled as human, it was rated with higher trust scores than AI-labeled information,} which is supported by Reis et al.~\cite{Reis_Moritz}, who found that people value human advice more when aware of AI's involvement, especially in health context. 
\major{This observation appears to generalize across domains, whereby an AI label can diminish people's perceived quality, even if the AI source was initially deemed superior.}
This includes AI art~\cite{Horton2023-ih}, general communication~\cite{ai_people_heard_label}, medical advice~\cite{Kerstan}.
\major{Even in clinical decision-making scenarios, people tend to prefer human decision-makers over AI, perceiving the latter as less dignified~\cite{Medical_AI_human_dignity},} further highlighting a deep-seated bias against AI involvement in sensitive health-related contexts.
\major{Moreover, Epstein et al.~\cite{Epstein2023} found that not only the presence of a label, but also its wording, can significantly affect trust. For example, people perceive content labeled as ``AI-assisted'' more favorably than ``AI-generated'', indicating that subtle linguistic framing influences users’ willingness to trust. This suggests that beyond binary source disclosure, the design and language of labeling also play a critical role in shaping perception.}

The qualitative findings (Sec \ref{sec:qual}) confirmed that participants expressed greater trust in human expertise, which they associate with verified knowledge, accountability, and human empathy. In contrast, they viewed the lack of consciousness, ethical judgment, and transparency in AI as diminishing their perceived trust.
The perspective expressed by our participants aligns with De Freitas et al.'s~\cite{factors_attitudes_AI} work about psychological factors affecting attitudes toward AI acceptance, \major{which identifies opacity (lack of transparency or explainability) and emotionlessness (absence of empathy or moral understanding) as key factors driving user resistance to AI tools.}
Our respondents echoed these concerns by highlighting AI’s lack of transparency and moral reasoning, especially in healthcare contexts, where trust is closely tied to perceived ethical awareness and human empathy. These reactions may also reflect a broader skepticism about machine consciousness~\cite{machine_consciousness}.

\major{These findings can be interpreted through the MATCH model~\cite{responsible_ai}, which conceptualizes trust in AI systems through three components. In our context, actual source of the information (human vs. LLM) corresponds to \textit{model attributes}, reflecting users’ judgments of competence and reliability. Disclosure labels (AI vs. human) act as \textit{afforded cues}, shaping trust perceptions independently of content quality. Participants’ perceptions, such as associating human expertise with trust or distrusting AI due to its lack of professionals, reflect \textit{trust heuristics}, where users rely on cognitive shortcuts in uncertain of complex health contexts. This framing emphasizes that trust is not only a response to information content but also to how the system communicates authorship and identity, and how users emotionally and cognitively process these trustworthiness cues~\cite{trust_automation} which was further explored through implicit behavioral and physiological responses in the following section.}

\major{Summarizing}, while AI is increasingly recognized for its competence, our findings underscore the role of transparency as a trustworthiness cue framed in the MATCH model~\cite{responsible_ai}, emphasizing the need for transparent AI-powered systems~\cite{Liao2023} and authentic information~\cite{ai_authenticity_1,transparent_ai} to build trust, particularly when providing nuanced health advice~\cite{Broom2005, Kerstan}.  
However, our study also cautions against over-reliance on labeling as a trust mechanism. As highlighted in prior work~\cite{scharowski_certification_2023}, labels can create a false sense of security and may inadvertently reinforce the ``implied truth effect''~\cite{Pennycook2020}, where unlabeled content is assumed to be accurate. 
These findings point to the need for more context-sensitive and dynamic approaches to communicating AI involvement in health information systems.


\subsection{Behavioral and Physiological Features Can Vary by Health Information Source and Label}

\major{Our results demonstrated that the effects of label and source are also evident at the behavioral and physiological level.}
Prior work has shown value in leveraging behavioral and psychophysiological sensing across fake news detection in social media~\cite{Abdrabou_Yasmeen} and information-seeking tasks~\cite{physio_search}, where such signals are indicative of visual attention and information processing in these tasks. 
With respect to trust, Ajenaghughrure et al.'s~\cite{Ajenaghughrure} review found that while \major{psychophysiological levels of trust perceptions (e.g., arousal) can be detected (e.g., using EEG or ECG), how such responses behave during user interactions (in real-time) remains underexplored.}
In the context of our study, we first explored \textbf{(RQ2)} whether such signals vary during health information processing across human versus AI-\revise{sourced} content, and essentially whether such signals can serve as a means of verifying and possibly predicting self-reported trust scores (Sec. \ref{ml_analysis}). 
We found that participants displayed distinct gaze patterns related to the source and labeling of the presented \revise{presented information}. Specifically, we found that longer fixation duration, higher fixation counts, and larger pupil dilation were associated with information labeled as human-generated, suggesting a deeper cognitive engagement with this human-labeled information, \major{suggestive of higher trust.}
Conversely, information labeled as AI-generated prompted more scanning behavior \major{(i.e., reflected in increased saccadic movements and shorter fixation durations),} indicative of increased verification processes. 
These results corroborate existing research from others (e.g., Just et al.~\cite{Just1980-tk} and Rayner et al.~\cite{Rayner1998-rc}) who likewise found that gaze patterns, especially the fixation and saccade behaviors, are indicative of cognitive processing and information verification \major{relevant to trust assessment and dynamics.}

For the peripheral physiological signals, while we found significant differences in features such as RMSSD and skin temperature when users encountered labeled health information, no such differences were found in skin conductance (SCL and SCR) measurements. 
It is worth speculating what this means: these indicators aligned with users' self-reports, where \revise{health information labeled as from AI elicited higher HRV (i.e., RMSSD) than the label of human professionals.}
\major{Higher HRV is typically associated with lower physiological arousal, possibly reflecting less cognitive processing or more relaxed state.}
\major{
This interpretation is consistent with the meta-analysis by Kim et al.~\cite{hrv_2}, which found that HRV reliably decreases under stress or increased cognitive demands, and increases under lower arousal or more comfortable conditions.
Indeed, HRV is one of the most commonly used psychophysiological indicators in trust research~\cite{Ajenaghughrure}, able to detect subtle variations in user state during human-computer interaction. 
Although Ajenaghughrure et al. caution that trust classification using physiological signals remains an open research challenge.
}
\major{Furthermore, the pattern of reduced physiological arousal in response to AI-labeled information aligns with the gaze data in our study, which suggested less attentional engagement (e.g., shorter fixations, fewer regressions) with AI-labeled content compared to human-labeled information. These findings suggest that participants may have processed AI-labeled health information with lower cognitive and emotional investment.}
Similarly, higher skin temperature levels were observed with both AI-labeled and LLM-sourced information, \revise{suggesting lower emotional arousal and stress levels, reflecting participants' psychological interpretation of trust~\cite{Ahmad_Muneeb}}.
I.e., participants gave higher trust scores to the LLM-sourced information compared to human-sourced, and showed lower physiological arousal with the AI labels than human labels.

\major{These behavioral and physiological responses deepen our interpretation of trust formation grounded in the MATCH model~\cite{responsible_ai}. While the online survey study (Study 1) focused on how users respond to model attributes and afforded cues (i.e., health information source and labels), we further extend the analysis to \textit{trust heuristics}, the implicit, affective processes that guide user trust-related judgments under uncertainty. Physiological responses like HRV and skin temperature likely reflect affective dimensions of trust (e.g., comfort, emotional arousal), whereas gaze patterns and fixation behavior index cognitive engagement. 
This layered interpretation aligns with calls to distinguish between cognitive and affective trust as investigated by Lee and See~\cite{trust_automation} which is grounded in the most widely used and accepted ABI trust model from Mayer, Davis, and Schoorman~\cite{organizational_trust_model}, suggesting that trust in AI-generated health content is not just explicitly reported but also embodied in users’ implicit affective reactions.}

\major{Taken together, these sensing signals could serve as a useful means to corroborate how users react and feel toward content perceived to be sourced from humans versus AI, while providing an additional layer of information about information processing and associated affect.}


\subsection{\major{Considerations: Toward Trust-Aware AI for Health Information Seeking}}

Our findings offer actionable design considerations for stakeholders designing or developing LLM-powered health information tools. 
These include interface designers and developers of adaptive AI systems. 
We outline practical considerations as below, grounded in the findings of this work.



\subsubsection{For UI Designers of Health Information Interfaces}

\major{\textit{Designing and placing labels for trustworthy interfaces.}}
As a key element of user interfaces, transparency labels play a crucial role in promoting trustworthy AI design~\cite{Liao2024}. 
Our findings show that labeling content as AI-generated consistently reduced trust compared to identical content labeled as human-generated. This suggests that while transparency is critical, poorly framed labels can inadvertently erode trust.
Given the critical role of UX for responsible and transparent AI design~\cite{Liao2024}, we find it important to foster trust already at the interface level when presenting health information.
\major{Prior work highlights the need for balance: too little transparency risks deception, while too much may undermine confidence~\cite{ai_transparency_trust}. Therefore, designers should carefully consider not only whether labels are present, but also how they are phrased and styled.
Research from Epstein et al.~\cite{Epstein2023} shows that both presence and framing can significantly shape user trust. 
Insights from privacy nutrition labels~\cite{nutrition_label} further demonstrate that visual choices of design, such as simplifying symbols, using color intensity to signal risks, and providing accessible visual explanations for technical terms, can improve users’ accuracy, efficiency, and satisfaction~\cite{nutrition_label}.}

\major{Our eye-tracking data supports this: participants gave more fixation counts to AI-labeled health information while more fixation counts to human labels. This indicates that labels strongly influence both user attention and trust judgments. Effective placement is therefore crucial: labels should appear in or near high-attention areas such as headlines or primary content zones, and be styled with moderate emphasis, visible, but not distracting.}

\major{Taken together, these insights point toward ``trust-aware'' UI design,} where transparency labels are not just added for compliance but are thoughtfully designed and positioned to foster trust without bias. Visual elements such as trust meters or engagement indicators could further reflect the health information system’s trust assessment and communicate how health systems interpret user interactions, making transparency both informative and supportive of trust.

\major{\textit{Uniform UI structure across health topics.}}
Our findings also showed that trust ratings did not vary across information types, suggesting that a uniform interface structure can be used across health content categories, allowing design efforts to focus more on trust-sensitive features like labels and source attribution rather than varying UI by topic.


\subsubsection{For Developers of Adaptive LLM-Powered Health Information Systems}

\major{\textit{Real-time user states estimation is feasible.}}
Our findings show that behavioral and physiological signals (e.g., fixation and pupil size) varied across conditions, showing potential in predicting self-reported trust and source attribution. 
\major{These results suggest the feasibility of integrating user-state modeling into adaptive health information systems, echoing recent efforts in Human-Computer Interaction that leverage physiological signals to guide interactive system design and development~\cite{Chiossi2024}.}
For instance, Boonprakong et al.~\cite{Boonprakong2023} develop bias-aware systems that use physiological data for cognitive load estimation~\cite{cognitive_load_physio}, and study~\cite{Ajenaghughrure_modeling} predicts trust using psychophysiological measures. 
Understanding users' implicit states has the potential to enable the health system to better support health information seeking, flag moments of confusion or disengagement, and ultimately improve trust perceptions in health information.

\major{\textit{Toward ``disclosure-aware'' interfaces.}} 
Building on this, our findings suggest the opportunity to build ``disclosure-aware'' health systems or interfaces that can dynamically adjust the transparency labels based on real-time user states. 
For example, when the system detects low attention (e.g., reduced fixations), it could highlight source labels to encourage more critical engagement. 
Conversely, when signs of cognitive overload or skepticism emerge (e.g., sustained focus on labels, increased pupil dilation), the system could simplify or temporarily de-emphasize the label to prevent unnecessary distrust, particularly when the content is accurate and clearly presented. 
Moreover, such ``disclosure-aware'' interfaces could provide on-demand explanations of labels, giving users deeper transparency only when users seek it. 

\major{This vision resonates with the concept of attentive user interfaces by Hummel et al.~\cite{adaptive_ui_gaze}, which sense and respond to users’ attentional states to ensure that key digital nudges are not overlooked. 
Extending this logic, transparency labels could be made on demand, surfacing prominently when attention is low, and simplifying when signs of overload or skepticism arise. 
Such attentional feedback loops point toward health information systems and interfaces that are not only disclosure-aware but also attention-adaptive, dynamically balancing clarity, trust, and cognitive load.}


Overall, this work advances HCI efforts to design AI health information systems and user interfaces that are not only transparent but also trust-aware and adaptive. 
By revealing how users respond to different information sources and disclosure labels, our findings offer actionable insights for both designers and developers.
These considerations can help calibrate trust more effectively, reducing over-reliance, mitigating undue skepticism, and ensuring that both AI- and human-generated health information are presented in ways that support informed judgment.


\subsection{Limitations and Future Work}

Our study had several limitations that should be considered when interpreting the findings.

First, while our findings suggest that behavioral and physiological signals show potential in reflecting trust-related responses, we caution against overinterpreting them as direct indicators of trust, a complex and subjective construct~\cite{Liu1,ti,Vereschak2024}. 
Such signals can be influenced by unrelated factors like attention, physical arousal, or contextual noise~\cite{Cacioppo}. 
Without careful contextualization, these signals could be misinterpreted as significant in scenarios where they merely represent contextual noise.
Future research should integrate additional modalities (e.g., fNIRS~\cite{Boonprakong2023}, EEG~\cite{Michalkova2024}) to more robustly capture underlying cognitive states. 
Moreover, translating these findings into real-world applications (e.g., web-based gaze tracking~\cite{Mounica2019}, rPPG from facial videos~\cite{McDuff2014RemoteMO}) raises ethical concerns regarding consent, data privacy, and potential over-reliance on AI~\cite{Ethical, ethics_1}. 
Hence, any deployment must adhere to legal regulations (e.g., European AI Act~\cite{EUAIAct}) and prioritize continuous consent based on on-device security and privacy controls.

Second, the controlled lab environment may have influenced participants’ responses, as being observed might heighten scrutiny of AI-labeled information, \major{potentially amplified by societal caution toward AI.} 
\major{However, such an a ``mere observer effect'', are likely just typical for controlled psychological experimental conditions, where participant awareness of observation can subtly affect behavior~\cite{mere-observer-effect}.}
While these settings are valuable for minimizing external confounders and ensuring reliable comparisons across conditions, future studies should nevertheless validate these findings in real-world environments to account for potential differences in naturalistic behaviors.

Third, our study measured trust at a single time point and relied on self-reports rather than actual decision-making actions. While this provides initial insights, trust is inherently dynamic and context-sensitive, often influencing real-world decision-making under uncertainty~\cite{Sillence}. 
Capturing only static trust ratings may miss how trust evolves over interactions or translates into behavior, such as whether individuals follow AI- vs. human-sourced advice.
\major{Future work should adopt longitudinal, action-oriented paradigms (e.g., Ecological Momentary Assessment~\cite{ema}) to better reflect how trust evolves over time and influences real-life health decisions. This would yield a more ecologically valid understanding of trust in LLM-powered health context.}

Fourth, while all LLM-generated responses were reviewed for consistency with human-authored content, \major{we did not explicitly screen for stylistic aspects such as tone, clarity, or writing uniformity,} which may influence perceived trust. 
Besides, this study focused on a single LLM (GPT-4o), and the findings may not generalize across other models (e.g., Claude, Gemini, Llama), which vary in output quality and style. 
Moreover, \major{we did not include an in-task manipulation check to assess whether participants consciously perceived the actual source behind the labeled information. However, we acknowledge that perceived source awareness could influence trust independently of disclosed labels.}
Future work should evaluate the role of stylistic linguistic features across different LLMs, with regard to trust in AI. 
Additionally, to better understand how users react to AI-generated content, future studies should incorporate perceived-source ratings (e.g., post-task questionnaires or detectability checks) to assess whether trust judgments are mediated by users’ ability to distinguish AI- from human-authored responses.

Lastly, our participant sample (notably WEIRD~\cite{WEIRD}) across both studies was not representative of the general population, further limiting generalization. 
This is particularly relevant for groups with varying levels of AI literacy or differing baseline trust in technology. Acknowledging this limitation helps specify to whom these findings most apply. Nevertheless, our study provides a key initial step toward understanding the impact of source and labeling in online health information. \revise{Future expansion to include participants from varied demographics can enhance our understanding of how trust in health information is perceived across different groups.}


\section{Conclusion}

Through a mixed-methods crowdsourcing survey (N=142) and within-subjects lab study (N=40), we found that AI-generated health information is trusted more than content sourced by human professionals, regardless of labeling, while human labels are trusted more than AI labels. 
Furthermore, we found that trust perceptions in personal health information are not only influenced by the source and label but also vary at behavioral and physiological levels. 
Our work highlighted the importance of considering AI transparency labels when measuring trust in online health information, and in developing techniques for verifying subjective trust perceptions and automatically inferring if and when to apply transparency labels based on sensed behavioral and physiological data. 
As such, we invite future research on understanding and designing for the physiology of online human-AI interactions, within and beyond AI-powered health information systems.


\newpage


\section*{Acknowledgments}
This work is funded by the European Commission in the Horizon H2020 scheme, awarded to Jos A. Bosch (TIMELY Grant agreement ID: 101017424). 

\section*{Disclosure Statement}
The authors declare no conflicts of interest related to this study.




\newpage


\bibliographystyle{ACM-Reference-Format}
\bibliography{main}


\end{document}